\pdfoutput=1
\documentclass[journal=jacsat,manuscript=article]{achemso}

\usepackage[version=3]{mhchem} % Formula subscripts using \ce{}
\usepackage[colorinlistoftodos,prependcaption,textsize=small, color =lime]{todonotes} % for notes
\usepackage{caption}
\usepackage{subcaption} % caption and subcaption necessary for subfigures
\usepackage[per-mode=power]{siunitx}[=v2] % version 3 not available on arxiv % cant use version 3 options: uncertainty-mode = full
\DeclareSIUnit\atomicmassunit{u}
\DeclareSIUnit\elementarycharge{\text{\ensuremath{e}}}
\DeclareSIUnit\U{U}
\DeclareSIUnit\Th{Th}
\newcommand\mz{\textit{m/z}}
\usepackage[colorlinks = true]{hyperref}

\graphicspath{{dependencies/}}

\author{Paul Fremdling}
\author{Tim K. Esser}
\author{Bodhisattwa Saha}
\affiliation[University of Oxford]
{Chemistry Research Laboratory, Department of Chemistry, University of Oxford, 12 Mansfield Road, Oxford OX1 3TA, UK}
\author{Alexander Makarov}
\affiliation{Thermo Fisher Scientific, Bremen, 28199, Germany}
\alsoaffiliation{Biomolecular Mass Spectrometry and Proteomics, Bijvoet Center for Biomolecular Research and Utrecht Institute for Pharmaceutical Sciences, University of Utrecht, Padualaan 8, 3584 CH Utrecht, The Netherlands}
\author{Kyle Fort}
\affiliation{Thermo Fisher Scientific, Bremen, 28199, Germany}
\author{Maria Reinhardt-Szyba}
\affiliation{Thermo Fisher Scientific, Bremen, 28199, Germany}
\author{Joseph Gault}
\altaffiliation{Vertex Pharmaceuticals, 86-88 Jubilee Avenue, Milton Park, Abingdon, OX14 4RW, UK}
\author{Stephan Rauschenbach}
\email{stephan.rauschenbach@chem.ox.ac.uk}
\affiliation[University of Oxford]
{Chemistry Research Laboratory, Department of Chemistry, University of Oxford, 12 Mansfield Road, Oxford OX1 3TA, UK}
\alsoaffiliation[MPI Stuttgart]
{Max Planck Institute for Solid State Research, Heisenbergstrasse 1, Stuttgart DE-70569, Germany}

\title[Preparative Mass Spectrometry]
 {A preparative mass spectrometer to deposit intact large
	native protein complexes}

\abbreviations{MS,prepMS,}
\keywords{preparative mass spectrometry, electrospray ion beam deposition}

\begin{document}
\singlespacing % requesterd by arXive
%%%%%%%%%%%%%%%%%%%%%%%%%%%%%%%%%%%%%%%%%%%%%%%%%%%%%%%%%%%%%%%%%%%%%
%% The manuscript does not need to include \maketitle, which is
%% executed automatically. The document should begin with an
%% abstract, if appropriate. If one is given and should not be, the
%% contents will be gobbled.
%%%%%%%%%%%%%%%%%%%%%%%%%%%%%%%%%%%%%%%%%%%%%%%%%%%%%%%%%%%%%%%%%%%%%
\begin{abstract}
Electrospray ion-beam deposition (ES-IBD) is a versatile tool to study structure and reactivity of molecules from small metal clusters to large protein assemblies. It brings molecules gently into the gas phase where they can be accurately manipulated and purified, followed by controlled deposition onto various substrates. In combination with imaging techniques, direct structural information of well-defined molecules can be obtained, which is essential to test and interpret results from indirect mass spectrometry techniques.

To date, ion-beam deposition experiments are limited to a small number of custom instruments worldwide, and there are no commercial alternatives. Here we present a module that adds ion-beam deposition capabilities to a popular commercial MS platform (Thermo Scientific\textsuperscript{TM} Q Exactive\textsuperscript{TM} UHMR mass spectrometer). This combination significantly reduces the overhead associated with custom instruments, while benefiting from established high performance and reliability.

We present current performance characteristics including beam intensity, landing-energy control, and deposition spot size for a broad range of molecules. In combination with atomic force microscopy (AFM) and transmission electron microscopy (TEM), we distinguish near-native from unfolded proteins and show retention of native shape of protein assemblies after dehydration and deposition. Further, we use an enzymatic assay to quantify activity of a non-covalent protein complex after deposition an a dry surface. Together, these results indicate a great potential of ES-IBD for applications in structural biology, but also outline the challenges that need to be solved for it to reach its full potential.
\end{abstract}
\clearpage

% \todo[inline]{Possible Journals: ACS nano, small, (Analytical Chemistry)}
\section{Introduction}
\label{sec:intro}

% Background: Sample preparation for microscopy
Cryogenic electron microscopy (cryo-EM), low-energy electron holography (LEEH) and scanning probe microscopy (SPM) are complimentary imaging techniques to probe the structure and conformation of biomolecules at sub-nm resolution \cite{renaud_cryo-em_2018,longchamp_imaging_2017,muller_atomic_2008,abb_carbohydrate_2019,wu_imaging_2020}.
Cryo-EM has evolved into a leading method for high-resolution imaging of biological macromolecules\cite{kuhlbrandt_resolution_2014,bai_how_2015,yip_atomic-resolution_2020}. LEEH is a low-energy electron, single-particle microscopy method that allows to image highly flexible proteins in their individual conformations\cite{ochner_low-energy_2021}. SPM reveals the connectivity of branched oligosaccharides \cite{wu_imaging_2020} and allows access to the electronic structure of individual molecules \cite{kahle_quantum_2012, kley_atomic-scale_2014}.

All three methods require samples produced at highest standard to work optimally.  LEEH and high-resolution SPM require ultra-pure, UHV-compatible substrate conditions and greatly profit from chemical purity of the adsorbate.\cite{longchamp_imaging_2017}. For cryo-EM, the preparation of homogeneous, high-quality samples can be challenging, especially for complex biomolecules.
%\todo{too specific, no direct link to cryo-EM?}Stretching or threading is limited to DNA \cite{payne_molecular_2013,zimmermann_dna_1994} 
Conventional sample preparation for cryo-EM proceeds through the plunge freezing method, which has been enormously successful, but can be time-consuming and resource-intensive, and homogeneity is limited by solution-based purification techniques.\cite{agard_chapter_2014,drulyte_approaches_2018,noble_routine_2018,chorev_use_2020}

% What is ESI-IBD?
Electrospray ion beam deposition (ES-IBD) is a preparative mass spectrometry\cite{cyriac_low-energy_2012,johnson_soft-_2016} technique, capable of producing highly purified molecular samples for single molecule imaging. It is routinely used for SPM with smaller (bio)molecules\cite{wu_imaging_2020,rauschenbach_electrospray_2006,hamann_ultrahigh_2011, rauschenbach_mass_2016, walz_navigate_2021, abb_two-dimensional_2016,deng_close_2012,rinke_active_2014} and has been demonstrated also for TEM\cite{vats_electron_2018,vats_catalyzing_2021, prabhakaran_rational_2016, mikhailov_mass-selective_2014}, LEEH\cite{ochner_low-energy_2021,longchamp_imaging_2017}, and recently cryo-EM \cite{esser_mass-selective_2021, westphall_3d_2021}. In contrast to organic molecular beam epitaxy (OMBE),\cite{mccray_mbe_2007,koma_molecular_1995} ES-IBD is not limited to small and volatile molecules. In ES-IBD, molecules are ionised in an electrospray ion source, transferred into the gas phase, and mass-analysed in vacuum. Then, the ion beam is mass-to-charge-ratio filtered and deposited with a controlled landing energy onto a suitable substrate. ES-IBD is often referred to as "soft landing" at lower collision energies, or "reactive landing" at higher collision energies or if the collision results in formation of a  covalent bond to the surface. It enables new reaction pathways\cite{krumbein_fast_2021,yang_anionanion_2021} and surface modifications \cite{su_design_2019}.

% Requirement: beam control
In addition to the requirements for ESI mass spectrometry, ES-IBD needs an intense ion beam\cite{pauly_hydrodynamically_2014,rauschenbach_electrospray_2006,su_multiplexing_2021, bernier_transfer_2020} with well-defined energy distribution, to enable fast sample preparation with controlled landing energy. The width of the beam-energy distribution is crucial, as it defines the collision energy distribution and limits the landing energy range. Reported values for full width at half maximum (FWHM) range from 2 to \SI{10}{\electronvolt} per charge\cite{krumbein_fast_2021,rauschenbach_electrospray_2006,walz_compact_2020,hamann_electrospray_2011}. The beam-energy width determines the minimal landing energy, which can be used without deflecting a significant portion of the ion beam. Narrow beam-energy distributions enable controlled exploration of shallow conformation spaces \cite{anggara_exploring_2020}. Low landing energy is particularly important for highly-charged protein complexes, as their absolute landing energy is proportional to their charge state.

Likewise, the beam intensity determines the deposition time for a given deposition area and particle density. 1 pAh ($3.6 \times 10^{-9}$ Coulomb or 22 billion charges) is the charge deposited by a 1 pA ion beam over one hour. In practice, \SIrange{5}{20}{pAh} are sufficient for imaging\cite{esser_mass-selective_2021,rinke_active_2014,longchamp_imaging_2017,ochner_low-energy_2021}. This charge allows to deposit on a several \si{\mm^{2}} large sample with a sub-monolayer coverage that enables imaging of isolated particles. Beam currents of more than \SI{20}{\pA} ensure typical deposition times of less than half an hour, so multiple deposition conditions can be tested in a day. However, precise mass-selection inherently reduces the current available for deposition, since all ions except the selected one are removed from the beam. Finally, an accurate current measurement on the level of \SI{1}{\pA} is needed to achieve reproducible coverage.

% Requirement: native
For sample preparation of biological macromolecules, the structural integrity of fragile biomolecules has to be maintained for the entire ES-IBD process. Native MS retains covalent and most non-covalent interactions within a protein complex\cite{bakhtiari_protein_2019,sharon_mass_2007,tamara_high-resolution_2021} and can be integrated to ES-IBD. Nevertheless, it remains unclear to which extent ionisation, liquid-gas-phase-vacuum transfer, and soft landing affect non-covalent interactions and hence the conformation and structure of the protein complexes.

% Relevance to science
Currently, the barrier to widespread use of ES-IBD is still high and there is no commercial instrument available. Academic instrument developers have designed preparative MS mainly for small and medium size molecule deposition\cite{franchetti_soft_1977,heiz_chemical_1997,miller_soft-landing_1997,laskin_soft-landing_2008,hamann_electrospray_2011,rauschenbach_mass_2016,su_multiplexing_2021} and only few of these instruments can handle native proteins complexes.\cite{longchamp_imaging_2017,ochner_low-energy_2021,mikhailov_mass-selective_2014} To be universally useful for molecular ion-beam deposition, ES-IBD instruments need to be good mass spectrometers, and have a high beam current in addition to the features needed for beam control and deposition.

Commercial, analytical instruments typically are excellent mass spectrometers, but have insufficient beam intensity for ES-IBD and lack the flexibility in design and software to integrate deposition as an additional workflow. As a minimum requirement, a native ES-IBD/MS must handle large, low-mass-to-charge-ratio protein ions with a molecular weight of up to a megadalton. Whilst some home-built or converted machines can do this \cite{benesch_separating_2010,longchamp_imaging_2017,walz_compact_2020,westphall_3d_2021,ochner_low-energy_2021}, their mass filter, collisional activation or beam control is severely restricted in comparison to commercial instruments.

Here, we show how to convert a proven, commercial, analytical, native MS to a native ES-IBD platform. It has an intense, well-controlled ion beam, which we characterise with current and energy measurements. Three different methods are used to demonstrate that the platform is suitable for near-native deposition: Protein heights observed in SPM images show globular features when preparing samples using native ES-IBD, compared to denatured, conventional MS conditions. Using TEM, we demonstrate the importance of landing energy control to preserve near-native structural features. Finally, we show that a non-covalent enzymatic complex retains activity after ES-IBD.

\section{Results and Discussion}

\subsection{Instrument setup and modification}

\begin{figure}[h!]
 \includegraphics[width=\linewidth]{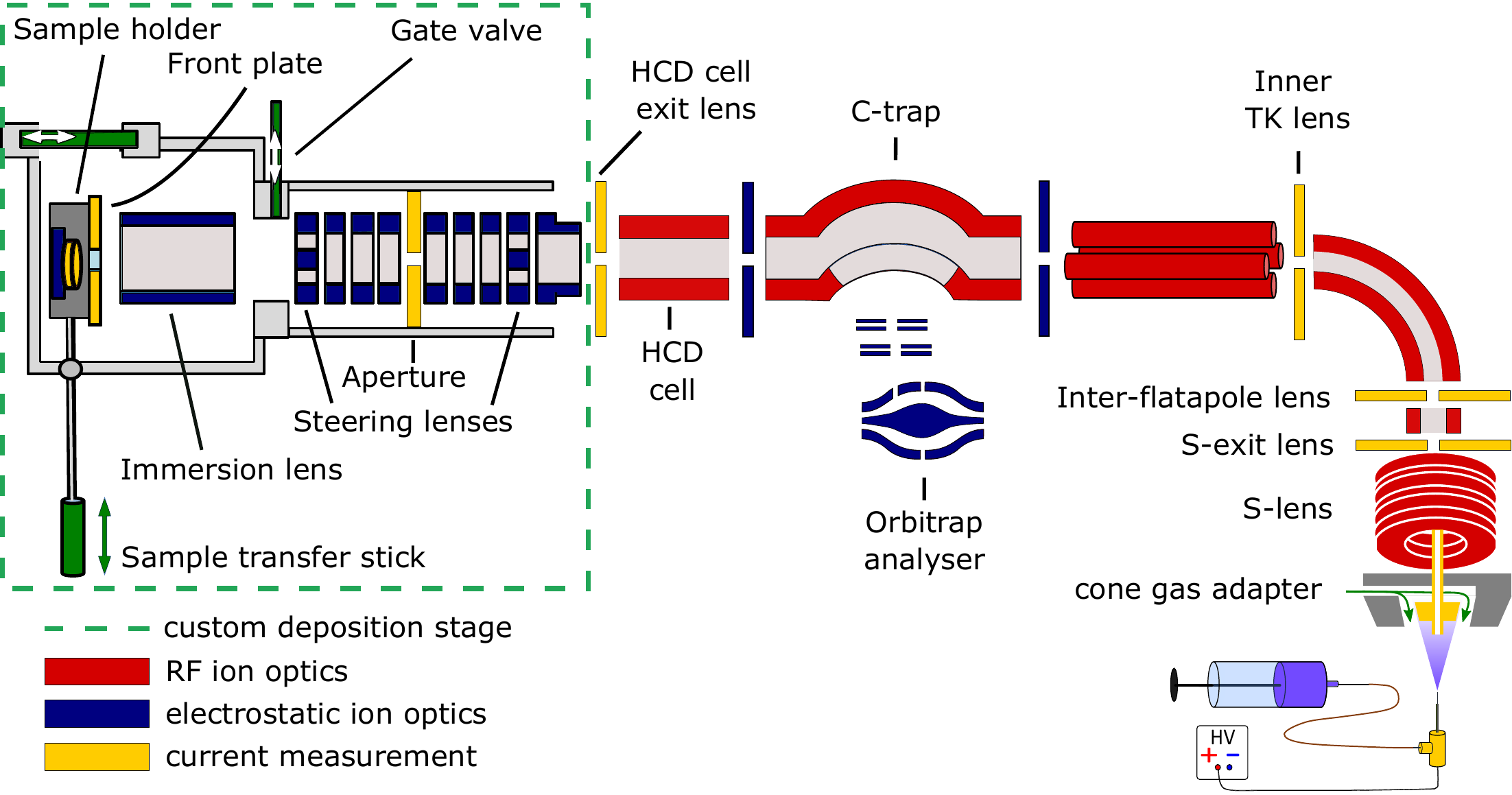}
 \caption{Schematic view of the Q Exactive UHMR mass spectrometer modified for deposition. Custom landing stage to deposit two microscopy samples and measure energy on the left. UHMR with improved source for better transmission on the right.}
 \label{fgr:instrument}
\end{figure}

We have converted a Q Exactive UHMR instrument (Thermo Fisher Scientific, Bremen, Germany) into a preparative mass spectrometer by adding a custom-built landing stage downstream of the Higher Energy Collisional Dissociation (HCD) cell. Fig.~\ref{fgr:instrument} shows a scheme of the instrument. The added stage contains electrostatic lenses to focus and steer the beam onto a sample holder, containing two sample positions and a retarding grid energy detector (this scheme only shows a single sample in the sample holder).

A sample transfer rod moves the samples in and out of the deposition chamber. That process takes two minutes including pumping and venting.
To monitor the beam intensity, the ion current is measured at the landing stage and on apertures throughout the instrument, which was modified to add this capability (yellow elements in Fig.~\ref{fgr:instrument}).
In addition, we have increased the S-exit lens diameter from 1.4 mm to 2.5 mm and added a custom cone gas adapter to increase transmission efficiency and thus achieve shorter deposition times (see Methods).

\subsection{Deposition workflow}

First, we load up to two samples, typically TEM grids or highly oriented pyrolytic graphite (HOPG) substrates, into the sample holder and insert it into the deposition stage. We create an ion beam and check the composition with the Orbitrap mass analyser and set the quadrupole mass filter to select the species required for deposition.

To optimise the beam intensity for deposition, we switch to beam mode. In this mode, the C-trap and the HCD cell guide the ions in a continuous beam, instead of intermittently pulsing the beam into the Orbitrap mass analyser. All direct current (DC) potentials within the Q Exactive UHMR instrument were kept at default values, which minimize activation during transmission from source to the deposition stage (see Fig.~\ref{fgr:UHMR_landing_potentials_SI}a). This usually means that potential gradients are as low as possible especially in regions where collisions with the background gas occur.

Next, the beam is steered onto the energy detector. In front of the collector plate that is used to measure the ion current, the detector has a metal grid to apply retarding voltages.
% If the electric potential between collector plate and grid is higher than the kinetic energy of the ion, it cannot reach the collector.
Ions with a total energy below their potential energy at the grid cannot reach the detector plate.
Hence, we record the ion current at the detector plate as a function of the grid potential to obtain the beam energy.

The difference between the beam energy and the retarding sample potential determines the landing energy. We typically use a range from \SIrange{2}{100}{\electronvolt} per charge depending on the specific application. For deposition, we finally steer the beam onto the sample and start integrating the detected sample current, to measure when the desired coverage is achieved. During deposition the beam composition is checked periodically using the mass analyser.

\subsection{Beam-energy distribution}

The total energy of the ion beam and its distribution are pivotal parameters for the ES-IBD process because they define the collision energy with the surface. The total energy distribution is determined by the potential along the beam path and the interactions of the ions with the background gas. Hence, it can be influenced by the local pressure, which is a function of pumping speed and shape of the vacuum vessel, and by the applied radio-frequency (RF) and DC voltages.

In our instrument, total energy is measured via the retarding grid detector integrated in the sample holder (see Fig.~\ref{fgr:instrument} and Methods). Ions moving through the upstream part of the instrument experience small gradients in pressure varying from \SI{0.01}{mbar} in the HCD cell to high vacuum in the landing stage. While keeping all other conditions constant, we can obtain intense beams with different sets of voltages applied to the electrodes of the ion optics along the beam path. We investigated the influence of two distinct sets of potentials on the beam-energy distribution, one with higher and one with lower potential gradients (see Fig.~\ref{fgr:UHMR_landing_potentials_SI}b).

For this investigation, we used an ion-beam of denatured and a native bovine serum albumin (BSA). Denatured BSA yields a wide range of charge states between +44 and +15 (\SIrange{1600}{4500}{\Th}, Fig.~\ref{fgr:MS_BSA_SI}a). The native BSA beam contains the monomer as well as undefined, higher-order aggregates. Their mass-to-charge ratio is 3900 (+17, monomer) to 10200 Th (aggregate, Fig.~\ref{fgr:MS_BSA_SI}b). After the C-trap, the ions pass through the HCD cell and the electrostatic lenses and finally reach the energy detector.

At the detector we were able to measure the beam's intensity and total energy ($E_\mathrm{tot}$)
\begin{equation}
	E_\mathrm{tot} = E_\mathrm{kin} + E_\mathrm{pot}.
\end{equation}
The ions kinetic energy, $E_\mathrm{kin}$, only depends on velocity. Its potential energy, $E_\mathrm{pot}$, depends on charge state and position in the electric potential landscape. The reference for $E_\mathrm{pot}$ and $E_\mathrm{tot}$ is electrical ground by convention. Hence, an ion with a negative $E_\mathrm{tot}$ moving towards a grounded electrode would not reach it, because once all kinetic energy is converted $E_\mathrm{pot} = E_\mathrm{tot} < 0$.

\begin{figure}
 \includegraphics[width=0.8\linewidth]{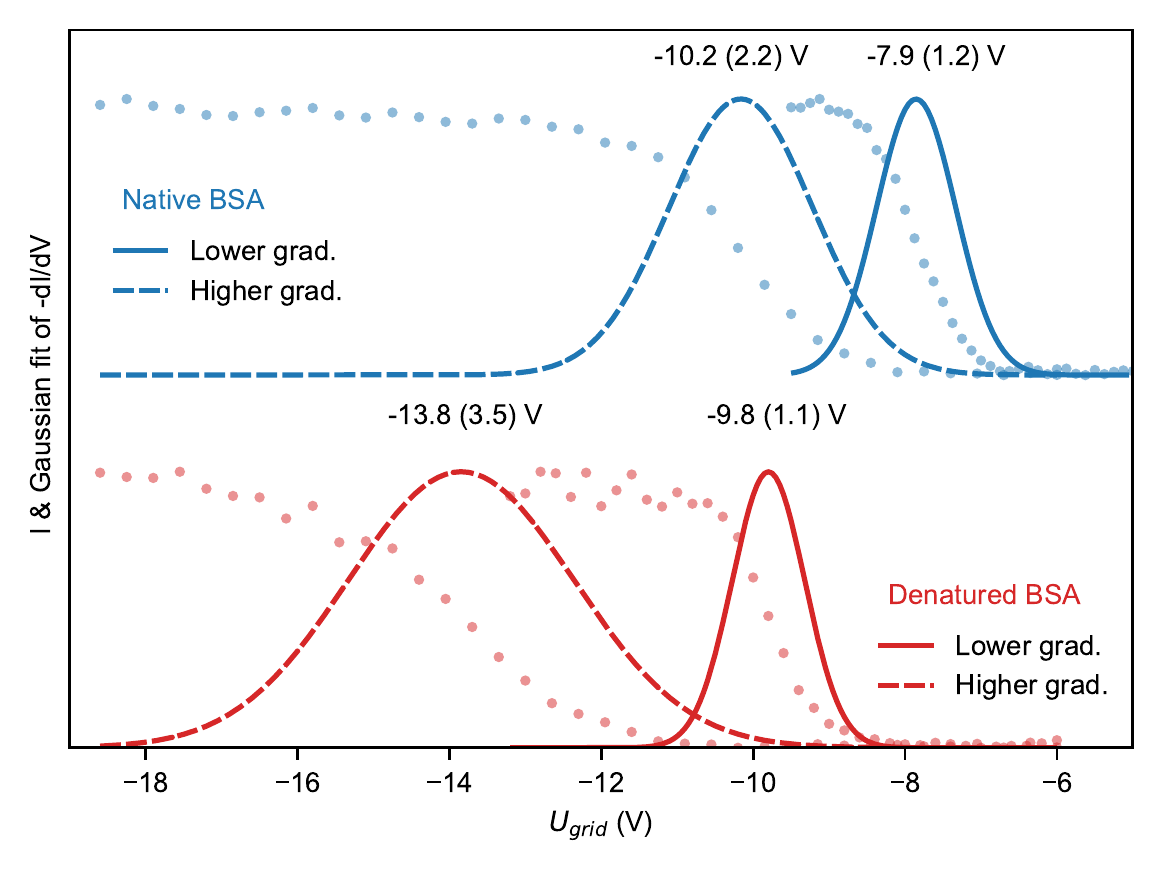}
 \caption{Beam-energy distribution measured for denatured and native BSA ion beams for two different potential gradient settings as shown in Fig.~\ref{fgr:UHMR_landing_potentials_SI}b. Dots: Ion current measured as function of the retarding grid potential $I(U_\mathrm{bias})$. Lines: Gaussian fit for first derivative $\mathrm{d}I/\mathrm{d}U_\mathrm{bias}$, corresponding to the total beam energy per charge ($E_\mathrm{tot}$) in eV per charge. FWHM is given in parentheses.}
 \label{fgr:energy}
\end{figure}

Fig.~\ref{fgr:energy} shows beam-energy distributions measured under different conditions. They are represented as Gaussian fits to the first derivative of the beam current, $I$, with respect to the grid bias, $U_\mathrm{grid}$. Based on the above conventions, the grid potential $U_\text{grid}$ corresponds to total energies. Clearly, the state of the ion, folded or unfolded, as well as the chosen potential landscape influence the energy mean value and energy width, in the following given as $E(\Delta E)$.

When a lower DC gradient for focusing within the electrostatic lens was applied, the denatured BSA $E_\mathrm{tot}$ was
$-9.8(1.1)$ \SI{}{\electronvolt} per charge.
It is lower by \SI{4}{\electronvolt} per charge and widens by \SI{2.4}{\electronvolt} per charge when choosing a higher gradient instead.
Native BSA's $E_{tot}$ follows a similar trend, albeit with a higher $E_\mathrm{tot}$ mean of \SI{-7.9}{\electronvolt} per charge with the lower gradient
%(+ \SI{1.9}{\electronvolt} per charge)
and \SI{-10.2}{\electronvolt} per charge for the higher gradient.
%(+ \SI{3.6}{\electronvolt} per charge).

%Discussion
The interplay between local pressure and ion acceleration in the electrostatic lens determines $E_\mathrm{tot}$. Ions thermalise in the HCD cell to a total energy of \SI{-5}{\electronvolt} per charge, which is defined by the axial DC-potential. From there they enter the electrostatic lenses. Although the pressure rapidly decreases, the ion's mean free path is significantly shorter than the distance between the HCD exit lens and the next aperture and hence energetic ion-background gas collisions will occur. The ions gain kinetic energy ($E_\mathrm{kin}$) between two collisions proportional to the DC gradient (electric field, see figure \ref{fgr:UHMR_landing_potentials_SI}b) along the flight path in the landing stage. The relative loss of kinetic energy per collision depends mainly on the mass of the collision partners, with the absolute loss per collision higher at higher $E_{kin}$. The randomness of the impact angle between gas and ion causes a distribution in energy loss, which is wider for high $E_{kin}$. Thus, a high potential gradient causes a large decrease in $E_\mathrm{tot}$ and widens $\Delta E_{tot}$, the width of the distribution (see SI).

Two factors explain the lower $E_{tot}$ for the denatured protein. First, the number of collisions in the electrostatic lens increases with the unfolded protein's larger collisional cross section\cite{douglas_collisional_1992}. Second, the denatured protein ions' higher charge states raise the overall $E_{kin}$ (for the same value of energy per charge), which leads to higher energy loss in collisions as compared to the low charge state, native ion.

In summary, when transferring an ion beam from high-pressure RF optics into high vacuum, magnitude and distribution of $E_\mathrm{tot}$ is a function of the DC gradient, background pressure, ion charge, and collision cross section (CCS). For a given type of ion, efficient pumping and a weak DC gradient ensure a narrow distribution of total beam energy, enabling all ions to land on a substrate downstream with a similar collision energy.

Here, using low gradients, the $E_\mathrm{tot}$ distribution (FWHM $\leq \SI{1.2}{\electronvolt}$ per charge) is sharper than previously reported literature values (FWHM $\geq \SI{2.2}{\electronvolt}$ per charge) \cite{krumbein_fast_2021,rauschenbach_electrospray_2006,walz_navigate_2021,hamann_electrospray_2011}, pointing to gentle conditions in which gas-phase activation is minimal. Given that the lower gradient conditions also achieved high transmission and good beam focus, and we retained them for all other experiments presented here.

\subsection{Transmission}

\begin{figure}
 \includegraphics[width=\linewidth]{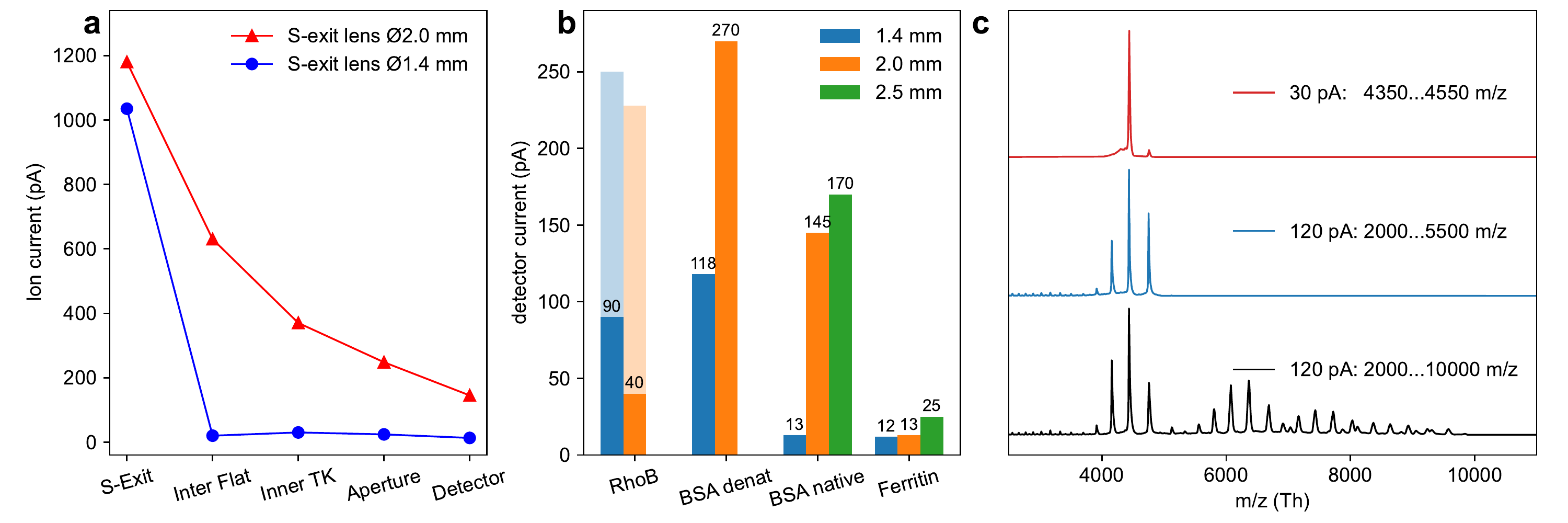}
 \caption{Transmission properties.
 \textbf{a} Ion current across the instrument before and after increasing the S-exit-lens diameter, measured at different ion optics.
 \textbf{b} Typical ion currents at the energy detector for different S-lens diameters. (Values equivalent to sample currents). Protein ion currents increase with aperture size. RhoB currents don't follow the trend, due to a defocusing effect. Currents on preceding optical element are shown in light colours.
 \textbf{c} Native BSA current on energy detector decreases with the narrowing width of the mass filter window. }
 \label{fgr:transmission}
\end{figure}

High transmission is crucial for deposition experiments, since the particle flux directly determines the deposition time for a given coverage and sample surface area. Using a typical concentration of \SI{3}{\micro\mole\per\liter} and assuming a \SI{1}{\ul\per\hour} nano electrospray flow rate with \SI{100}{\percent} ionisation efficiency, a \SI{1.2}{\nA} emission current of native BSA ($z=15$) would be generated (see SI for details). However, under these conditions we measured initially only \SI{13}{\pA} at the sample position in the Q Exactive UHMR instrument with an unmodified source region. An initial measurement indicated a 1 nA current in the first vacuum chamber (Fig.~\ref{fgr:transmission}a). This may include ionised solvent and contaminants. There was also a sharp drop in current between the S exit lens and the inter flatapole lens.

To improve the transmission performance, we enlarged the inner diameter of the S-exit lens stepwise from 1.4 to 2.0 and finally to \SI{2.5}{mm}. With the \SI{2.5}{mm} opening, the ion current at the sample for large, native proteins doubled to \SI{25}{\pA} and for medium sized, native proteins the current grew more than ten-fold to \SI{170}{\pA} (Fig.~\ref{fgr:transmission}b). There was no measurable effect for Rhodamine B (RhoB), a relatively small ion with an \mz{} of 443 Th. All currents reported here are routinely reached with fluctuations of up to \SI{80}{\percent}, due to emitter performance.

The overall transmission is further affected by mass-filtering, where a narrow \mz-window not only suppresses contamination, but can also reduce the flux of desired analyte molecules.
Fig.~\ref{fgr:transmission}c illustrates how the width of the mass-filter window affects the native BSA current: Removing higher-order agglomerates has little effect on the sample current (bottom to mid panel). In this case, it was possible to filter a single charge state whilst retaining a third of the total current.

% Discussion
In contrast to the protein ion currents, RhoB current does not change with increasing S-lens diameters. Likely, a different beam profile as compared to heavy protein ions causes this behaviour. Thanks to its low \mz, RhoB experiences a stronger effective potential than high \mz{} protein ions within the S-lens. Thus, it can remain closer to the optical axis, reducing losses at the transfer apertures.

The modifications to increase the ion current are vital for depositing larger molecules. They allow to test several deposition conditions on a single experiment day, where particle densities of \SI{3000}{\per\micro\m\squared} or more are needed for efficient cryo-EM or SPM. For our applications, this is usually achieved with a deposited charge of \SI{15}{\pA h}. The modifications lead to a deposition time of approximately \SI{0.5}{\hour} for large native protein complexes.

Whilst necessary for preparative MS, our modifications cause the gas flow into the injection flatapole collision cell to become significantly higher. The pressure in the flatapole rises as a consequence and could decrease the in-source-trapping effectiveness.

%SR_done

% \clearpage
\subsection{Ion-Beam Shape and Control}
\label{ssec:spotsizetext}

\begin{figure}
 \includegraphics[width=.9\textwidth]{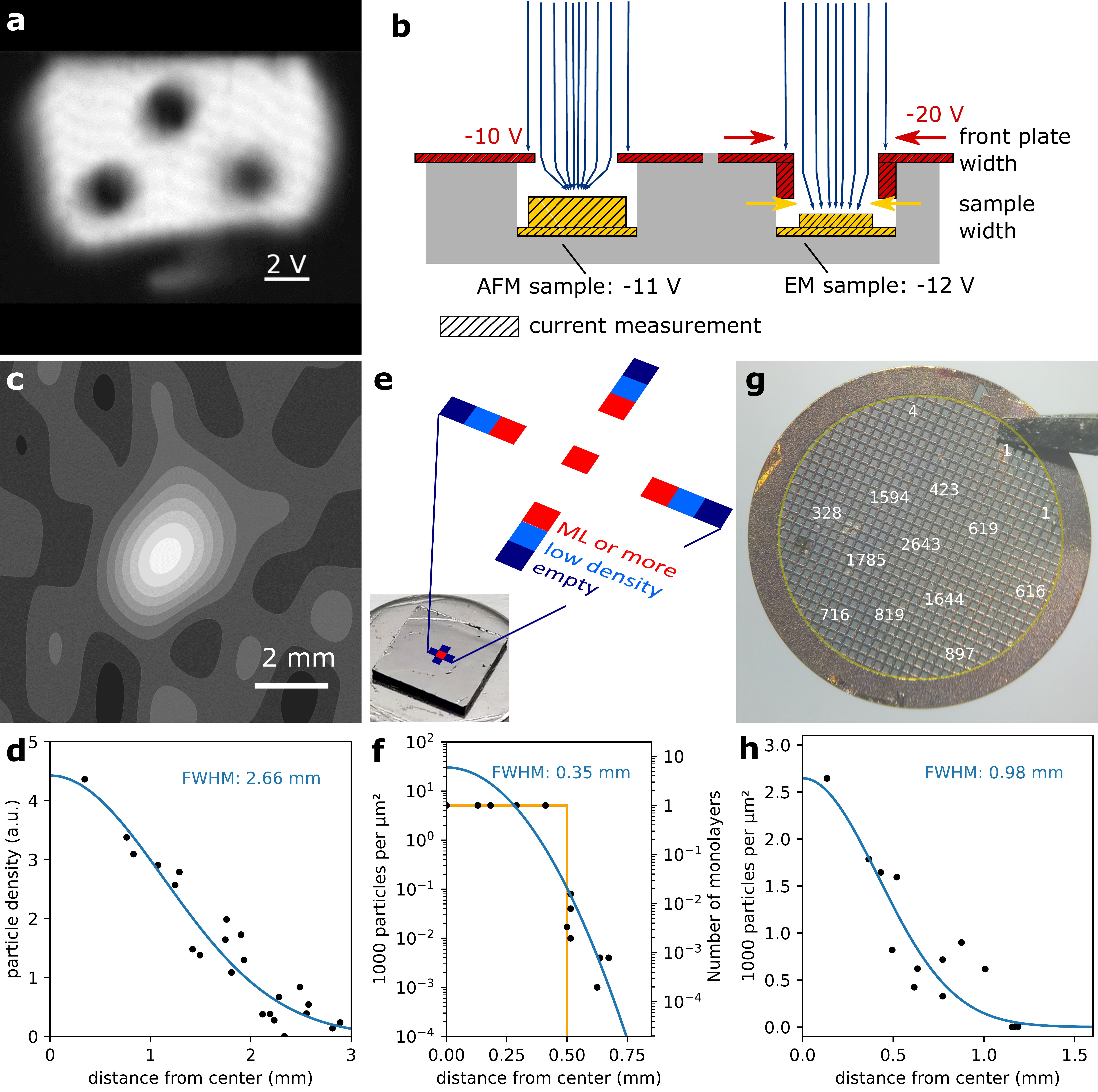}
 \caption{ Ion Beam Shape Analysis:
 \textbf{a} Ion-beam image of the sample holder front plate.
 \textbf{b} Sample holder section view: Different voltages influence focusing.
 \textbf{c} Ion-beam shape from deconvolution of \textbf{a}.
 \textbf{d} Data points and Gaussian fit of ion-beam intensity distribution.
 \textbf{e} HOPG sample used for AFM measurements and protein density distribution in the screened area.
 \textbf{f} Measured protein distribution on \textbf{e} (black dots), Gaussian fit (blue), and single monolayer model (orange).
 \textbf{g} Amorphous carbon grid used for TEM measurements and protein density distribution in particles per \textmu m$^2$.
 \textbf{h} Gaussian fit (blue) of density distribution in \textbf{g}.
 Individual AFM and TEM micrographs are shown in Fig.~\ref{fgr:spotsize_SI} in the SI.
 }
 \label{fgr:spotsize}
\end{figure}

% introductory paragraph
The ability to create a narrowly focused beam is essential to reduce the time needed to achieve the optimal particle density for SPM or TEM.
We used three different methods to assess the ion beam profile under typical experimental conditions.

% Front plate beam profile
First, we took an ion-beam image of the front plate of our sample holder (see Fig.~\ref{fgr:spotsize}a). For this, we scanned the beam with the deflection elements in the electrostatic lenses and recorded the current on the front plate. The resulting current image is a convolution of the front plate geometry and the beam shape. Deconvolution revealed a Gaussian-like beam profile (shown in Fig.~\ref{fgr:spotsize}c). A Gaussian fit gives a FWHM of 2.7 mm, only slightly larger than the diameter of the preceding aperture of 2 mm, which the beam typically passes without losses. The observed widening between the last aperture and the front plate is a consequence of the beam-energy distribution and the DC gradient in this section. A weak DC gradient moves the ions slowly in axial direction and gives them more time to expand radially. The beam profile obtained in this way is the profile at the front plate, whereas the samples are located a few mm behind and can be biased at a different potential.

%Transition
The beam profile is different on the sample, because the potential gradient between front plate and sample can focus the beam (Fig.~\ref{fgr:spotsize}b). We used AFM to determine protein density distribution after ion beam deposition on HOPG and TEM imaging after deposition on a TEM grid.

% AFM spot size
We typically use 5 mm wide HOPG chips (see Fig.~\ref{fgr:spotsize}e) as substrates for AFM imaging. For the example given here, we deposited 12.5 pAh of GroEL. Multiple AFM images were taken on the graphite sample, distributed along the length and width of the sample. We found that, for the specific DC potentials used in this experiment, most of the surface area was empty and proteins were localized in a small spot near the centre.
Surprisingly, we observed a transition from a clean, empty surface to a coverage of more than a monolayer within 250~$\mu m$.
We estimate the total number of GroEL particles as \SI{4.2E9}, from the deposited charge and average charge state of +67.

Because AFM cannot distinguish between single and multiple monolayer coverage, we can only roughly approximate the particle distribution. We fitted our data to two alternative models.
A Gaussian fit combines the total particle number with the particle density in the sub-monolayer coverage area. It suggests a deposition spot FWHM of just 350 \textmu m and a coverage of up to six monolayers at the centre. However, it fails to reproduce the sharp increase in density at the spot's boundary. Alternatively, we assume a monolayer density in the centre (ca. 5000 particles per \textmu m$^2$) with a sharp drop to 0 at .5 mm from the spot centre (orange curve in Fig.~\ref{fgr:spotsize}f). This model overestimates the density at the spot boundary. The real distribution is likely found between these two estimates.
As changing position on the sample can be tedious in AFM, other methods with wider field of view or faster change of position would be more appropriate to analyze particle distributions.

% TEM spot size
Thus, as a third approach, we deposited an apo/holo-ferritin mixture on a TEM grid covered with 3 nm amorphous carbon film (see Fig.~\ref{fgr:spotsize}g) and acquired micrographs at room temperature. The density of holoferritin iron cores was quantified on different grid squares. The resulting distribution is shown in Fig.~\ref{fgr:spotsize}h, together with Gaussian fit. A clear decrease of protein density from the centreed maximum to the edges of the grid is observed. The fit gives a FWHM of 1 mm and a total particle count of \SI{2.9E9}{}. We can compare this number to the estimate from the total accumulated deposition current of 20 pAh. Using the most abundant apoferritin charge state of +50, this corresponds to \SI{9.0E9}{} particles. We attribute the deviation partially to ambiguity of the charge state, due to the continuous mass to charge distribution of ferritin, caused by the randomness of the mass of the iron cores. Hence, the charge state distribution cannot be measured with ensemble MS techniques. This makes the calculation of the number of landed particles less accurate. In addition, apoferritin, which accounts for 40\% of the total ion-beam intensity, was not detected due to radiation damage.

% comparison of results
The different approaches to the measurements of the deposition spot size provide comparable results and show that the ion beam can be focused to reduce the preparation time of high-density protein samples. Differences in the spot size can be understood by the use of two different proteins, DC potentials, and different sample geometry.
% used -10 V on FP for AFM and -20 V for TEM (-10 V TEM sample failed)
The AFM sample is thicker, and thus closer to the front plate. This changes the local electric fields and leads to a different focus. We have observed that the deposition spot size can be tuned most effectively using the DC potential between front plate and sample.
The beam can also be defocused to create a more homogeneous distribution across the entire sample. Generally, either full monolayer coverage or few isolated particles can be achieved to optimize the sample for various imaging applications.

The size and shape of the deposition spot measured here is consistent with other observations. Secondary ion mass spectrometry together with infrared reflection absorption spectroscopy showed similar distributions of below- and above-monolayer coverage\cite{laskin_soft-landing_2008}. Most importantly, the strong influence of the fields directly at the sample suggest that more effective focusing could be achieved with dedicated ion optics installed at this location.
% stored for 2 weeks

%_SR_done

% \clearpage
\subsection{Control of conformation after landing by mass filtering and solution composition}

\begin{figure}
 \includegraphics[width=\linewidth]{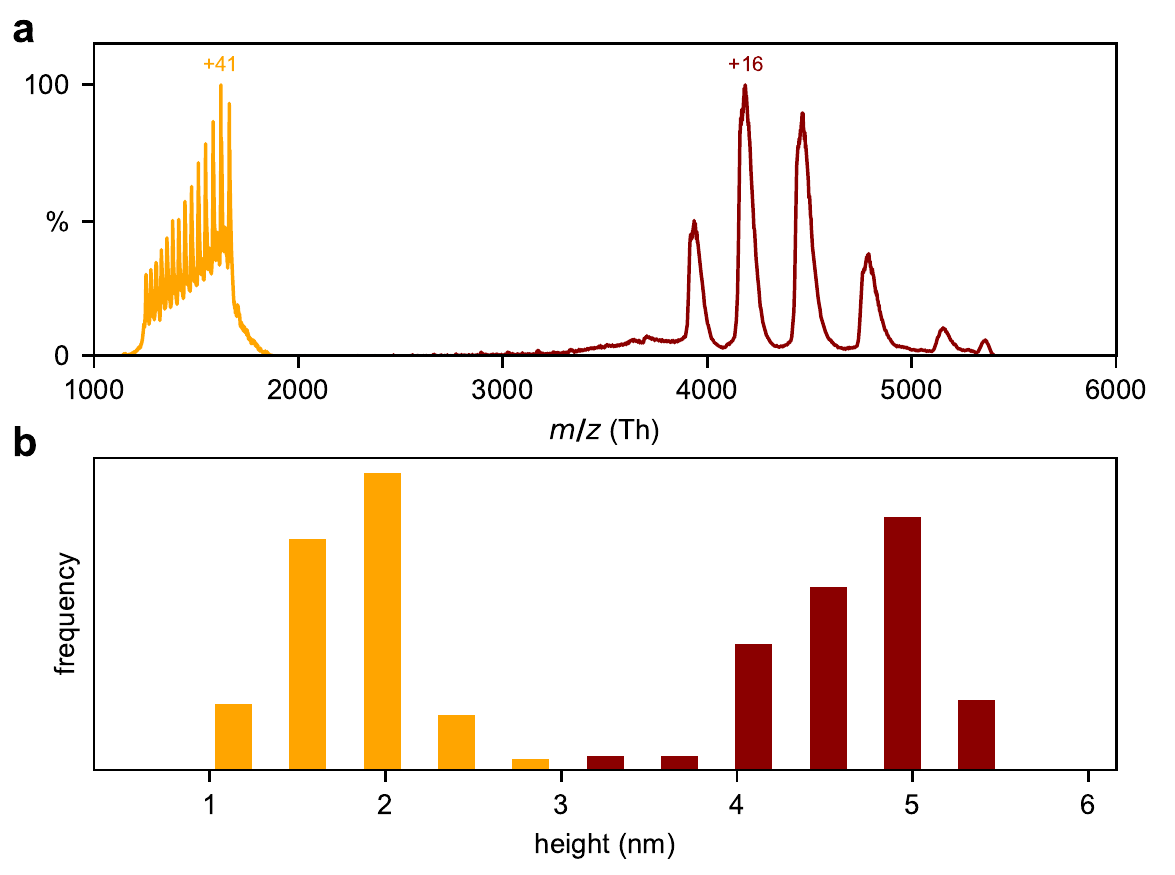}
 \caption{Native (red) and denatured (yellow) mass spectra and height histograms. Spray solution composition: Native \SI{200}{\milli\mole} \ce{NH4Ac}, denatured 73:24:3 \ce{MeOH}:\ce{H2O}:\ce{HCOOH}. \textbf{a} Mass spectra for native (filter window 2000...5000 m/z) and denatured (1250...1700 m/z) BSA. \textbf{b} Resulting height distribution measured with AFM after soft-landing on HOPG. N native = 47, N denatured = 60.}
 \label{fgr:QMS-height}
\end{figure}
% \sisetup{uncertainty-mode = separate} % to format stddev as +- % cant use version 3 options
\sisetup{separate-uncertainty = true} % to format stddev as +- % use version 2 options

It is established, for example by ion mobility spectrometry, that the three-dimensional (3D) conformation of proteins can be retained to a large degree in native ESI.\cite{shelimov_protein_1997} To study if such a native-like conformation can be retained in our instrument, we soft-landed BSA on HOPG using different solutions and instrument settings.

Fig.~\ref{fgr:QMS-height} shows two mass spectra of BSA. When using a solvent containing \SI{73}{\percent} \ce{MeOH}, \SI{3}{\percent} \ce{HCOOH} (formic acid), and \SI{24}{\percent} water and a conventional ESI source, high charge states were observed indicating that the protein is denatured and unfolded. We selected the charge states +40 to +53 with the mass filter for deposition. For a \SI{200}{\milli\mole\per\liter} \ce{NH4Ac} solution nano-sprayed at \SI{1.2}{kV}, much lower charge states between +14 and +17 are observed which indicate folded BSA. We selected only the BSA monomer for deposition.

After deposition, AFM images are taken and quantitatively analysed (see methods) to extract the height distribution, which allow an approximation of the shape of the adsorbed proteins. The height distribution is \SI{1.8\pm0.3}{\nm} for denatured BSA, and \SI{4.7\pm0.4}{\nm} for native BSA, given as mean $\pm$ standard deviation.

Adsorbates originating from highly charged, denatured protein ions appear much flatter than their low-charged native counterparts. This difference in height is consistent with proteins in completely unfolded and globular conformations, respectively. However, it is not possible to directly image the conformation of individual soft-landed proteins in ambient AFM. Firstly, the individual BSA molecules have undergone diffusion limited aggregation\cite{zhang_atomistic_1997} on step edges and terraces. Hence, the individual proteins cannot be identified unambiguously (Fig.~\ref{fgr:native_HOPG_SI} and \ref{fgr:denatured_HOPG_SI}). Secondly, the AFM radius of the tip is too large to resolve the lateral shape of the aggregates. Instead, a convolution of the tip shape and adsorbate shape is measured, but the height is reproduced with great accuracy ($<1$\r{A}).

%Discussion
This result proofs that the ionisation conditions, notably source and solvent, control the conformation of the soft-landed protein on HOPG. The CCS describes the ion conformation in the gas phase ahead of the landing event. The CCS of BSA measured in \ce{N2} for charge state +40 to +53 is \SIrange{134}{144}{\nm^2} \cite{elliott_simultaneous_2017}, for native BSA (+14 \ldots +17) it is \SI{45}{\nm^2} \cite{bush_collision_2010}.
Our measured heights are in good agreement with these values because high CCS, extended denatured conformations yield flatter agglomerates than native, compact ones. Therefore, protein height measurements after soft-landing can reveal pre-landing gas-phase conformations on mass spectrometers without IMS capability. This is consistent with previous observations that conformations are retained, on the level of a general shape, after soft-landing on a relatively inert surface like graphite.\cite{siuzdak_mass_1996,rauschenbach_electrospray_2006,mikhailov_mass-selective_2014}

%SR_done_

% \clearpage
% \sisetup{uncertainty-mode = full}
\sisetup{separate-uncertainty = false} % need version 2 options

\subsection{Mass-selective preparation of cryo-EM protein samples}
% introductory paragraph
For large, folded protein assemblies, cryo-EM has become one of the leading methods for structural characterization at atomic resolution.\cite{kuhlbrandt_resolution_2014,bai_how_2015} Negative-stain EM, on the other hand, is commonly used to screen sample quality before preparation of cryo-EM samples.
% relevance
Native ES-IBD has the potential to complement and accelerate established cryo-EM sample preparation workflows by selective sample preparation and direct correlation between cryo-EM density maps with complementary information about native interactions and small ligands from mass spectrometry. 
%In addition, precise control of the landing energy and substrate opens up a novel route to analyse composition, binding energy, and mechanical properties by imaging collision-energy specific states or fragments.

% Ion-beam deposition has already been employed for the sample preparation of various high-resolution imaging techniques, including SPM, TEM, and LEEH demonstrating a great versatility in terms of molecular size and properties.\cite{rauschenbach_mass_2016,abb2016,deng2012,longchamp_imaging_2017,rinke2014,vats2018,wu_imaging_2020,mikhailov2014} % redundant
Our ion-beam deposition instrument can cover TEM grids with mass-selected protein assemblies, with unprecedented landing energy control, for imaging in negative-stain EM and cryo-EM.
Native gas-phase protein ions are generated via native electrospray ionization, then mass selected, and deposited on TEM grids at room temperature. Grids are retrieved via the vacuum load-lock, transferred under ambient conditions and either stained using uranyl acetate or manually frozen in liquid nitrogen to create cryo-EM compatible samples while circumventing vitrification.

\begin{figure}
 \includegraphics[width=.9\textwidth]{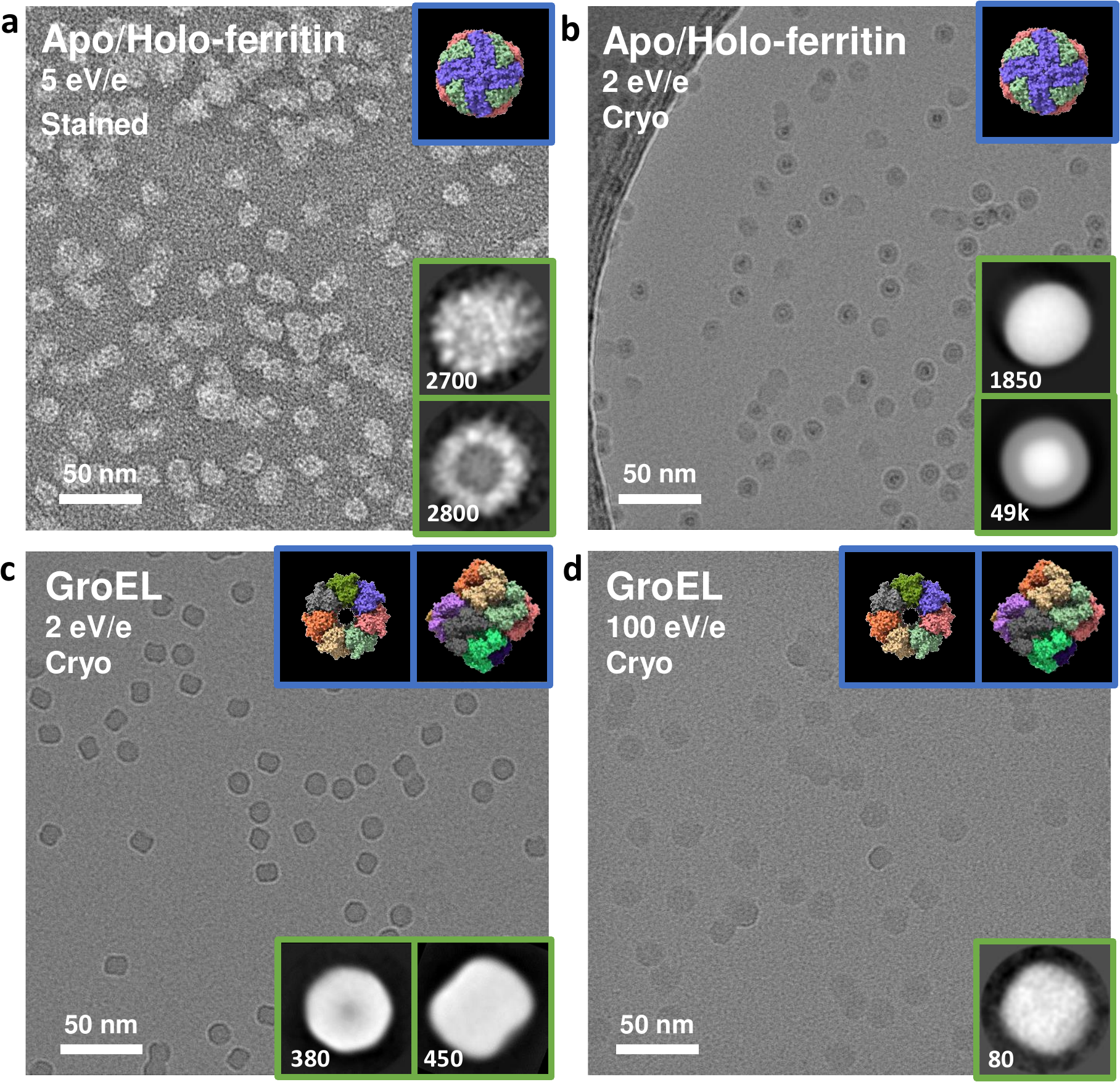}
 \caption{Negative-stain and cryo-EM micrographs of apo/holo-ferritin and GroEL after gas-phase purification and gentle deposition on TEM grids.
 \textbf{a} Apo/holo-ferritin, landing energy of 5 eV per charge, 30 nm amorphous carbon film, stained with uranyl acetate.
 \textbf{b} Apo/holo-ferritin, landing energy of 2 eV per charge, 3 nm amorphous carbon film, plunge-frozen in liquid nitrogen.
 \textbf{c, d} GroEL, landing energy of 2 eV, 100 eV per charge, 3 nm amorphous carbon film, plunge-frozen in liquid nitrogen.
 The insets show 3D models from the PDB (blue), rendered with ChimeraX\cite{pettersen_ucsf_2021} using PDB entries 7A6A for apoferritin and 5W0S for GroEL, and 2D classes of native ES-IBD samples (green) obtained using RELION 3.1. The number of particles in the 2D classes is given in the insets.
 }
 \label{fgr:TEM}
\end{figure}

Fig.~\ref{fgr:TEM} shows negative-stain and cryo-EM micrographs from native ES-IBD samples of apo/holo-ferritin (479 kDa) and GroEL (803 kDa). 3D models from the PDB (blue) and two-dimensional (2D) classes (green) obtained from single particle analysis in RELION 3.1 are shown as inserts.

% negative stain
In the micrograph of a negative-stain sample of an apo/holo-ferritin mixture, Fig.~\ref{fgr:TEM}a, individual proteins with and without iron cores can be identified.
The edges of the protein shell in the 2D classes are less defined than for a control sample made by conventional liquid deposition (shown in Fig.~\ref{fgr:TEM_SI}). The apo-ferritin 2D class indicates structural heterogeneity, likely due to a deformation of the hollow protein shell, while the holo-ferritin is stabilized by the presence of the iron core in its centre.

A similar workflow has recently achieved significantly higher quality for stained samples of GroEL, by landing in a glycerol matrix before negative staining, even without precise landing energy control.\cite{westphall_3d_2021} This highlights that landing, interaction with the solid substrate, and vacuum exposure can influence the structure of protein complexes, and a high level of control is needed to minimize deviation from native structures.

Combining ES-IBD of protein complexes with negative stain TEM, with or without liquid matrix, has great potential for screening applications. However, we have focused on cryo-EM sample preparation because negative staining ultimately limits access to high-resolution and information on internal structure.

% Cryo vs. stain
A micrograph of a native ES-IBD cryo-EM sample of the same apo/holo-ferritin mixture is shown in Fig.~\ref{fgr:TEM}b. The particles have a significantly higher contrast compared to conventional cryo-EM micrographs, due to the use of a 3 nm thin amorphous carbon film and the absence of ice. The ferritin protein shells are clearly visible around the iron cores and demonstrate conservation of protein complex topology. A slight deformation of the apoferritin is still observed, but it is smaller than for the stained sample, and the 2D classes show sharp rather than diffuse edges.

This result indicates that the deformation observed in Fig.~\ref{fgr:TEM}a is not only due to the deposition on dry samples at room temperature, but due to the influence by negative staining. We suspect that the exposure to the air-water interface in the staining step limits sample quality in this workflow.

% gentle vs. harsh landing energy
Finally we compare ES-IBD samples of GroEL prepared with landing energies of 2 and 100 eV per charge, imaged by cryo-EM and shown in Fig.~\ref{fgr:TEM}c and Fig.~\ref{fgr:TEM}d, respectively.
Top and side projections of GroEL can be identified unambiguously in the sample prepared at the lower landing energy. The features of the characteristic barrel shape, including the central cavity and heptameric symmetry in the top view, are already apparent in the single particle images. Particle dimensions indicate no lateral deviation from literature values. However, further detailed substructure, as observed in samples prepared by plunge-freezing, is not visible, which is attributed to small random changes in secondary and ternary structure, which blurs the images classes (see \citeauthor{esser_mass-selective_2021} for a detailed discussion).

In the sample prepared using a landing energy of 100 eV per charge, Fig.~\ref{fgr:TEM}d, individual particles are still clearly visible, but they are up to 30\,\% larger in diameter, and the distinctive structural features have disappeared. Identification of side and top views is no longer unambiguous. This clearly shows plastic deformation of the GroEL complex due to the energetic impact on the surface, as all other conditions were kept identical. Our workflow enables systematic investigation of the landing energy dependence of this deformation to infer mechanical properties of proteins and protein assemblies.

%_SR_done_

\subsection{Retention of enzymatic activity}

The difference in structural detail observed between the plunge-frozen cryo-EM samples and ES-IBD samples suggest a level of structural change. To study to what degree this structural change can affect the biological function of proteins, we tested whether the non-covalent protein complex ADH retains enzymatic activity after deposition. So far, this has only been shown for recalcitrant single-stranded proteins with no prosthetic groups such as trypsin \cite{ouyang_preparing_2003,volny_preparative_2005}.
%Trypsin, lysozyme, yeast hexokinase, bovine protein kinase A catalytic subunit landed on liquid glycerol:fructose:\ce{H2O} 65:35:5 surface \cite{ouyang_preparing_2003} and trypsin landed on plasma-treated metal \cite{volny_preparative_2005}
We adapted a photometric assay to quantify ADH activity by NADH production after landing on a surface.

\begin{figure}
 \includegraphics[width=\linewidth]{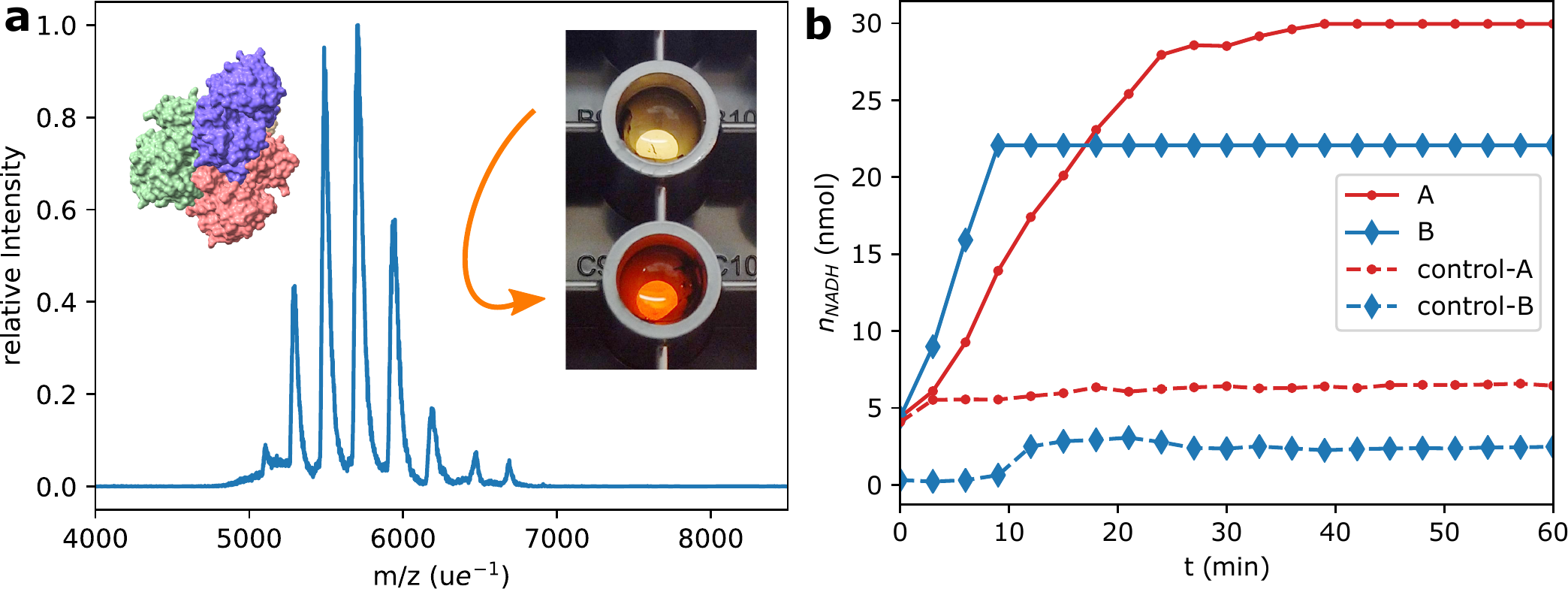}
 \caption{\textbf{a} Mass spectra of deposited, mass-selected ADH tetramer. Inset: Colour change from yellow to orange indicates an active ADH in the lower well. The black objects on the well's walls are the submerged ADH-coated conductive tapes.
 \textbf{b} Production of NADH by ADH after ES-IBD. The broken lines stagnating at the offset level are background controls, so the NADH production is specific for ADH activity. The absorbance measurement causes two separate artificial saturation levels due to different calibrations.}
 \label{fgr:ADH_assay}
\end{figure}

%Successful experiment with sumbersion
We deposited ADH on conductive carbon tapes with \SI{27}{\ng} (\SI{128}{\pA h}) ADH for repetition A, and for repetition B with \SI{22}{\ng} (\SI{102}{\pA h}). For each experiment 2 samples were made. Assuming a \SI{2.5}{\mm} diameter deposition spot, this corresponds to 2 monolayers on average. Fig.~\ref{fgr:ADH_assay} shows production of NADH by the samples together with background control submerged conductive carbon tapes. The ADH activity is proportional to the slope in of the curves in Fig \ref{fgr:ADH_assay}b. It was \SI{1.2}{mU} (A) and \SI{1.9}{\milli\U} (B). Minimal (A) or no (B) background activity was recorded in the corresponding time frame. The recovery, based on ADH data sheet activity (\SI{300}{\milli\U\per\g}) was \SI{14}{\percent} (A) and \SI{29}{\percent} (B). When the activity of the spray solution is taken as a reference (A: \SI{88}{\milli\U\per\g}, B: \SI{138}{\milli\U\per\g}), we find activities of \SI{48}{\percent} (A) and \SI{65}{\percent} (B) for soft-landed ADH. The positive control activity was lower than spray solution activity (A: \SI{56}{\milli\U\per\g}, B: \SI{117}{\milli\U\per\g}). We measured no activity for a \SI{27}{\ng} (\SI{128}{\pA h}) conductive carbon tape after 3 days storage in vacuum (Fig.~\ref{fgr:storage_SI}). (For further details on attempted ADH extraction refer to SI.)
% \open{We could add here that other substrates did not work, which makes a valid point in the discussion: could be due to different surface properties (collision: unlikely, polymer will be similar like amorphous carbon. adsorption: maybe, we need to dissolve them again. optics: plate reader compatibility.)}

%discussion
These results offer compelling evidence that a large, non-covalent protein complex can survive the entire ES-IBD workflow including ionisation, de-hydration, transfer into high vacuum, soft-landing, and re-solvation.
It is difficult to quantify the exact proportion of intact enzyme. Instead of the numerical value, the order-of-magnitude of the activity is relevant. A number of experimental uncertainties cause this: When reconstituting the commercially obtained, crystalline ADH, it is not known which proportion of the enzyme refolds incorrectly and remains inactive. We measured the spray solution concentration photometrically using a calculated attenuation coefficient. Surprisingly, a much higher proportion of deposited ADH than expected from these references was found to be active. Thus, we used an extrapolation and later a non-linear calibration (see methods).

Additionally, the conductive carbon tape could have blocked a small part of the plate reader beam path inside the well and increased absorbance. To mitigate errors, deposited ADH quantity should be cut to a third to remain in the linear range and the reading frequency increased.
The loss of all activity after 3 days storage in vacuum at room temperature might be a consequence of degradation, surface interaction or desolvation.
Further experiments are required to investigate if the soft-landed reconstituted ADH was the intact homo-tetramer. TEM images of ADH, soft landed under comparable conditions, indicate no fragmentation or change in quaternary structure.\cite{esser_mass-selective_2021}

%\clearpage

\section{Conclusions}

While the existing ES-IBD prototype instruments show some of the desirable features, a high-resolution, native, mass spectrometer with ES-IBD capability, which includes high beam intensity, ion beam monitoring and control, and adjustable, low and narrow deposition energy is currently not commercially available.

This work details the conversion of a high-performance serial Orbitrap mass spectrometer, designed for the very high mass range typical for native protein complexes, into an instrument for molecular ion beam deposition.
Beyond additional ion optics and a deposition stage, this requires the complete understanding of the instruments' beam handling in order to align the new components with the duty cycle of the original instrument. Also, obtaining sufficient intensity is a major achievement, for which small modifications to the existing ion optics were needed in addition to the implementation of ion current monitoring. Finally, an intuitive beam guiding and monitoring software is most helpful in characterising the performance and obtaining reliable, reproducible deposition results.

The focus of this instrument modification is the deposition and imaging of native proteins in order to add chemical selectivity to the protein structure determination process. This deposition and imaging work-flow is in development. Currently, electron density maps from samples prepared with ES-IBD lack the necessary resolution to determine if the protein complex structure has been completely preserved.\cite{westphall_3d_2021, esser_mass-selective_2021}. An alternative approach to check the integrity of the deposited protein is an enzymatic assay. It indicated that the activity of the non-covalent protein complex ADH was retained post-deposition.

The instrument developed here shows that a commercial platform can indeed be modified to perform deposition experiments reliably and under full control, while the excellent performance of the mass spectrometer is retained. We have not yet utilized the full capabilities for beam modification such as ion activation, or high-resolution selection of a fragment ion, but in principle these capabilities are available. Generally, the pulsed operation of the Orbitrap instrument allows for a manifold of operation modes, in which deposition can be integrated as part of the duty cycle.

% \clearpage

\section{Methods}
\subsection{Mass-filtered electrospray-ion-beam-deposition machine design}

We converted a Thermo Scientific{\textsuperscript{TM}} Q Exactive UHMR into a preparative mass spectrometer (Fig.~\ref{fgr:instrument}). The electrometer at the end of the HCD cell was removed to make space for a custom deposition stage. Analytical tandem MS still works unaffected in the modified UHMR.

The deposition stage contains a 2x8 element electrostatic lens to focus the ion beam. Steering lenses deflect the beam laterally to any position on the sample holder. A \SI{2}{mm} diameter aperture separates the two lens stacks. The first lens stack is pumped via the Q Exactive UHMR quadrupole. A \SI{67}{\liter\per\s} turbo pump in the deposition part pumps the second part (HiPace 80, Pfeiffer Vacuum GmbH, Asslar, EU). A CF 40 gate valve (series 01, VAT Vakuumventile AG, Haag, Switzerland) decouples the deposition stage from the analytical mass spectrometer. After the gate valve, an immersion lens shields the ion path from the electric potential of the grounded vacuum chamber. Hence, beams with negative Total Energy ($E_\mathrm{tot}$) vs. GND can pass.

The sample holder has two sample positions for EM grids or AFM samples and an energy detector to measure beam $E_\mathrm{tot}$. A custom sample transfer stick moves it from a load lock to high vacuum (HV). RBD 9103 HV floating picoampmeters (RBD Instruments Inc., Bend, USA) measure ion current on aperture, sample holder front plate, samples, and the energy detector. An ECH 244 crate with 2x EBS 180 $\pm$ \SI{500}{V} bipolar power supply insets control all DC voltages to deposition stage (ISEG Spezialelektronik GmbH, Radeberg, EU). Home-written control software for the picoampmeters and power supplies facilitates the ES-IBD workflow. It supports rapid 2D ion beam imaging, $E_{tot}$ beam measurement and automatic beam focusing optimisation.

To use sweep gas with the nano-ESI source, we milled a \SI{20}{\mm} bore in the cone gas adaptor. The S-lens diameter was increased from \SI{1.4}{\mm} to \SI{2.0}{\mm} and later to \SI{2.5}{\mm} to improve ion transmission. Consequently, gas throughput at the source turbo pump (Splitflow 310, Pfeiffer Vakuum GmbH, Asslar, EU) rose from approximately \SIrange{2.7}{5}{\milli\bar\liter\per\s}. % injection flatapole pressure from 0.1 mbar to 0.17 mbar (all approx., calculated from pressure gauge in Splitflow 310 fore vacuum hose)
We separated the fore pump system to protect the Splitflow 310. The S-lens chamber remained pumped by the factory-fitted Sogevac SV65BiFc fore pump (Atlas Copco, Stockholm, EU), and the Splitflow 310 was connected to an Edwards XDS 35i (Atlas Copco, Stockholm, EU) fore pump.

\subsection{Deposition workflow}
%Sample loading
The first step is to load two EM-grid or AFM highly oriented pyrolytic graphite (HOPG) targets into the sample holder. The transfer rod moves them from the ambient load lock to the high vacuum deposition chamber. Whilst the pressure therein decreases, we prepare the ion beam.\newline
%beam preparation
For native proteins, we use gold-coated \SI{1.2}{mm} glass capillary emitters. We select the minimum possible pressure to push the spray solution to the tip. This maximises emitter lifetime. We start the instrument in normal analytical configuration to check if the emitter is working. We set the mass filter window, then switch to beam mode. Both samples are kept at a high, repulsive potential to avoid uncontrolled deposition. In beam mode, the C-trap and the HCD cell guide the ions without pulsing into the landing stage. All DC potentials within the Q Exactive UHMR instrument are at the default values to guarantee activation-free transmission from source to the deposition stage. In contrast, analytical native MS typically uses strong gradients, often in pulsed modes, to desolvate or dissociate protein complexes\cite{hernandez_determining_2007}. HCD gas flow is set to 7 to thermalise the ion beam in there.\newline
%The ion beam is steered on the electrostatic lens' aperture connected to a picoampmeter.
To optimise the current, we change the emitter distance, backing gas pressure and the cone gas flow. If the current is sufficient for deposition, we switch to analytical mode and acquire mass spectra of the ion beam. The instrument is set beam mode again and the beam steered on the energy detector. The detector has a metal grid in front of the collector plate used to measure current. If the electric potential on the metal grid is higher than the total beam energy, the ions cannot pass. Hence, we record the detector collector plate current as a function of the grid potential to obtain the beam energy.

%deposition
Then, we select the retarding potential on the sample. The difference between the beam energy and the retarding sample potential determines the landing energy, typically \SI{5}{\electronvolt} per charge. We deflect the beam on the sample and start the sample current integration. Once the charge reaches the defined value, the repulsive potential is re-applied. The beam composition is periodically controlled using the mass analyser, including every time we replace the nano spray emitter.
TEM imaging deposition procedure has been already described \cite{esser_mass-selective_2021}.

\subsection{Energy width}
We used a native and a denatured BSA beam. For preparations see below. All DC voltages within the Q Exactive UHMR instrument were at the default values. For both beams, we applied a weak or strong DC gradient in the landing stage optics. This focused them through the electrostatic lens on the energy detector. Fig.~\ref{fgr:UHMR_landing_potentials_SI}b shows the different voltages applied to produce a weak or strong gradient.

The voltage on the detector metal grid was swept in 40 voltage steps around the expected beam-energy value. For every voltage step, we recorded the average of 60 detector current measurements. This dampens arbitrary or short-term periodic current fluctuations.
The negative differential of the current by the voltage was fitted with a Gaussian distribution. The fit gives the mean beam energy and its FWHM.

\subsection{Transmission}
To measure ion current within the Q Exactive UHMR instrument, we added breakout cables. To this end, we separated the transfer capillary voltage supply from S-exit lens. Breakout cables were connected to the S-Exit lens, the inter-flatapole lens, the inner Turner-Kruger (TK) lens and the HCD exit lens. A modified cone gas cap adaptor supplies the transfer capillary voltage. Each breakout cable connects a RBD 9103 picoampmeter to a DC ion optic and the corresponding power supply on the Q Exactive UHMR DC supply board. \newline
For the current measurement in the Q Exactive UHMR, we set the DC optic (e.g. the S-exit lens) to an attractive potential and the following RF ion optics axis DC (e.g. the injection flatapole) to a repulsive potential. This ensures the entire beam is collected on the DC optic in question. All voltages are in table SI \ref{tab:SI_pot_current}. In the deposition stage, we deflected the beam instead on the aperture or energy detector. \newline
In the next step, we moved the emitter sidewards away from the transfer capillary to block the ion beam at a preceding element. The current offset was recorded and the emitter moved back in position. Then, we recorded the current. All values in Fig.~\ref{fgr:transmission} are offset corrected. \newline
We used the heated ESI source for Rhodamine B and denatured BSA solutions. The nano-ESI source was used for native Ferritin and native BSA.

\subsection{Ion beam shape analysis and control}
1. On Front Plate: We obtained a 2D image of the front plate with a denatured BSA beam. We chose denatured BSA, as it reproducibly provides an intense and stable ion beam, which allows to collect high-quality images. To obtain a scanned image, we deflected the beam horizontally and vertically with the steering lenses whilst recording the current on the front plate. A 41 x 65 pixel scan was obtained in \SI{34}{min}. The image dimensions where then converted from volts to millimetres by calibration with the actual front plate size. The image represents a convolution of the sharp front plate geometry, a function of only 0 and 1, and the ion beam profile, assumed to have Gaussian shape. We used a Python script to deconvolute. It employs a binary filter to create a sharp version of the image and then applies the convolution theorem to obtain the beam profile. Finally, we used a low pass filter to remove high frequency components, originating from the non periodic image boundary. \newline
% AFM sample id: AFM0010
2. On a HOPG AFM sample: We deposited 12.5 pAh of GroEL and used a NanoScope MultiMode AFM for imaging. GroEL was prepared as described in subsection "Spray solution preparation". For deposition, we followed the standard workflow. Except for the front plate voltage. It was at \SI{-10}{\V}, as close as possible to the beam energy of \SI{-7\pm1.6}{\electronvolt} per charge, to minimise the deposition spot size. We acquired multiple 5 x \SI{5}{\um^{2}} images on a raster around the deposition spot to further assess the protein distribution. We used the dimensions of the cantilever to raster across the surface and reconstruct a density map. We manually counted the number of aggregates in each image.
%TEM. TEM sample id: EM0050
3. On a TEM sample: \SI{20}{pAh} ferritin were deposited on an amorphous carbon TEM grid (AGS160-4, Agar Scientific, Stansted, Great Britain). The front plate was at \SI{-20}{\V} to the standard work-flow focus. We used a mixture of apoferritin and holoferritin to obtain high contrast. Under the given conditions only holoferritin iron cores are visible. TEM images were recorded using an FEI Talos 200c at room temperature and under the given conditions only the holoferritin iron cores are visible. A python script
%using opencv and skimage % should we cite those?
was used to count the number of particles on the TEM images. Measuring the current on the sample for mass-selected apoferritin and ferritin, the ratio between them was determined as 40:60 and the particle counts were corrected accordingly. The density was determined on multiple grid squares as the average of particle counts of three images, divided by the image area. The coordinates of the individual grid squares were obtained according to the grid square size on a 400 mesh TEM grid.

\subsection{Spray solution preparation}
We purchased rhodamine B (R6626-25G), bovine serum albumin (BSA, A0281-1G), equine spleen ferritin (F4503-25MG), GroEL (chaperonin 60, C7688-1MG) and baker's yeast alcohol dehydrogenase (A7011-15KU) from Sigma Aldrich (Darmstadt, EU). Ferritin and GroEL preparation has been already described \cite{esser_mass-selective_2021}. We dissolved rhodamine B in 80:20 \ce{H2O}:iPr to \SI{1E-4}{\mole\per\liter}. We made a denatured \SI{4E-6}{\mole\per\liter} BSA solution for AFM deposition in 73:23:3 MeOH:\ce{H2O}:HCOOH. For all other denatured BSA measurements, we used a \SI{3E-6}{\mole\per\liter} 100:100:1 ACN:\ce{H2O}:HCOOH solution.
We desalted native BSA and ADH twice with size-exclusion chromatography columns (P6, 7326222, Biorad, Hercules, USA). These were equilibrated with \SI{0.2}{\mole\per\liter} ammonium acetate (A2706-100ML, Sigma Aldrich). Resulting concentrations were 2 to \SI{5E-6}{\mole\per\liter}.
For all preparations, deionised water with $\rho >= \SI{18.2}{\Mohm.\m}$ filtered through \SI{0.22}{\um} was used. All other solvents were MS grade from changing suppliers.

\subsection{AFM analysis}
Prior to deposition, each highly oriented pyrolytic graphite chip (HOPG, MikroMasch, Sofia, EU) was cut into 5x\SI{5}{\mm} chunks and glued with leit-silver (09937, Sigma Aldrich) on an AFM stainless steel support. We used a multimode AFM (asmicro, Indianapolis, USA) with a Scout 350 silicon tip (Nunano, Bristol, Great Britain) in tapping mode at room temperature. The AFM images were further processed with Gwyddion. We used the graphite step-edges for height calibration. We selected the highest point of each protrusion as height measurement.

\subsection{TEM}
Ferritin (F4503-25MG) and GroEL (chaperonin 60, C7688-1MG) samples were purchased from Sigma Aldrich. Sample preparation was carried out using a standard native MS workflow, including exchange of buffer to volatile ammonium acetate, as described before.\cite{esser_mass-selective_2021}. All samples were imaged using a Talos Arctica 200 kV (Thermo Fisher Scientific), and images were processed using RELION 3.1, as described in \citeauthor{esser_mass-selective_2021}.
For staining, 30 nm amorphous carbon TEM grids (AGS160-4H, Agar Scientific) were plasma cleaned before deposition. After deposition, dry grids were placed on 25 $\mu$L of 2\% uranyl acetate, blotted, and left to dry. A control sample was prepared by applying 4 $\mu$L of 10 $\mu$M ferritin in PBS to the grid for 2 minutes, followed by blotting, washing and staining as described above.

\subsection{Retention of enzymatic activity}
The workflow we developed combines ES-IBD with an adapted photometric alcohol dehydrogenase detection kit (ab102533, Abcam, Cambridge, Great Britain).

Principle: ADH-catalysed oxidation of Propan-2-ol yields NADH and Propanone: \ce{NAD+ + Propan-2-ol <--> NADH + Propanone}. NADH reacts with a colorimetric probe to form a bright yellow complex analysed at $\lambda$ = 450 nm. Whilst the manufacturer does not specify the exact mechanism of the kit, it is most likely based on the WST-8 to WST-8 formazan reaction \cite{chamchoy_application_2019}.

Preparation: All micro-centrifuge tubes and pipette tips were normal PP. All solutions were shielded from direct light and kept on ice, except where mentioned. Each kit was reconstituted according to the manual \cite{noauthor_ab102533_2014}, divided into 4 aliquots and refrozen at \SI{-20}{\celsius}. On the day of the deposition, we thawed one kit aliquot and the desalted ADH spray solution. We prepared two positive control ADH solutions from crystalline ADH in the supplied buffer to theoretical in-well activities of \SI{4E-10}{\mole\per\minute} and \SI{4E-9}{\mole\per\minute}. Reaction mix and background control solutions were prepared as in the manual and kept at room temperature.
All solutions were prepared for a \SI{150}{\ul} total volume in well. This is made up of \SI{50}{\ul} active solution (buffer for blank, buffer for extraction or positive control) and \SI{100}{\ul} of either reaction mix (with substrate Propan-2-ol) or background control mix (no substrate).
We measured the ADH spray solution absorbance and determined the concentration with a calculated absorbance coefficient of \SI{195440}{\liter\per\mole\per\centi\m}. Based on this concentration we prepared spray solution positive controls with the same theoretical in-well activity as the other two positive controls.

Deposition: We cut a conductive carbon double-sided tape (EM-Tec CT6, 15-000406, Labtech, Heathfield, Great Britain) in half. We removed two thirds of the protective film on the back and glued it to a stainless-steel AFM support. The entire protective film on the top side was removed and the target installed in the sample holder. We prepared two targets per repetition, one for the sample and one for the background control. To minimise contamination, we immediately installed the sample holder in the deposition vacuum chamber.
The deposition followed the standard procedure. The nano spray needle was protected from direct light. We deposited two tapes with \SI{27}{\ng} (\SI{128}{\pA h}) ADH for both repetition (A) and (C). For repetition (B), we deposited two tapes with \SI{22}{\ng} (\SI{102}{\pA h}). The mass was determined based on the total deposited charge, most abundant charge state and molecular weight of ADH. In repetition (B), the deposited amount was lower due to low sample current. The landing energy was \SI{5}{\electronvolt} per charge.

Submersion and Measurement: The entire kit except for the reaction mix / background control solutions was reverse-pipetted in a 96 Corning 3881 non-binding surface half area well plate (Corning Inc., Corning, USA). We were doing this in parallel to deposition of the second ADH target. This minimises both the time the deposited ADH targets spend in high vacuum and the time they are exposed to the atmosphere. Repetition (C) targets were left for 3 days in the high vacuum deposition chamber.
Then, we put the two deposited tapes in a well with their empty side facing the wall and the centre optical path free. The wells were already filled with \SI{50}{\ul} assay buffer to avoid gluing the tapes to the well's wall.
We added reaction mix or background control mix and closed the plate with a transparent lid. A FLUOstar Omega plate reader (BMG LABTECH GmbH, Ortenberg, EU) incubated the sample at \SI{37}{\celsius} and read absorbance at \SI{450}{\nm} every \SI{3}{min} for \SI{2}{h}.

Data analysis: The initial slope of the NADH production (repetition (A): minute 3...15, (B): minute 0...6) was used for activity calculation. We subtracted background activity only if it was positive. Due to the high proportion of active ADH after deposition, we had to extrapolate the linear calibration for repetition (A) in the absorbance range from 1.6 to 2.5 (\SI{18}{nmol} NADH). To attenuate arising errors, we extended a non-linear calibration for repetition (B) to 2.7 (\SI{15}{nmol} NADH).
\pagebreak

%% classes. This should be given within the "acknowledgement"
\begin{acknowledgement}
We want to thank the Nanoscale Science Department at the Max-Planck-Institute for solid state reseach, in particular Artur Küster, for the CAD-construction and manufacturing of the deposition stage. We acknowledge support from Thermo Fisher Scientific who provided the UHMR mass spectrometer within the framework of a technology alliance partnership.
TE acknowledges funding from the European Union’s Horizon 2020 research and innovation programme under the Marie Sklodowska-Curie grant agreement No 883387.

Competing Interests:
M.R.S., K.L.F and A.A.M. are employees of Thermo Fisher Scientific, the company that commercializes Orbitrap-based mass analyzers.

\end{acknowledgement}

\clearpage
\section{SI}
\begin{suppinfo}
\setcounter{page}{1} % restart counting
\setcounter{figure}{0} % restart counting
\renewcommand{\thefigure}{S\arabic{figure}} % add SI infront of number
\renewcommand{\theHfigure}{Supplement.\thefigure} % to create unique labels for hyperref

%\esibdheader % macro with title, authors and affiliations defined on first use

\subsection{Beam energy}
\begin{figure}
 \includegraphics[width=.9\textwidth]{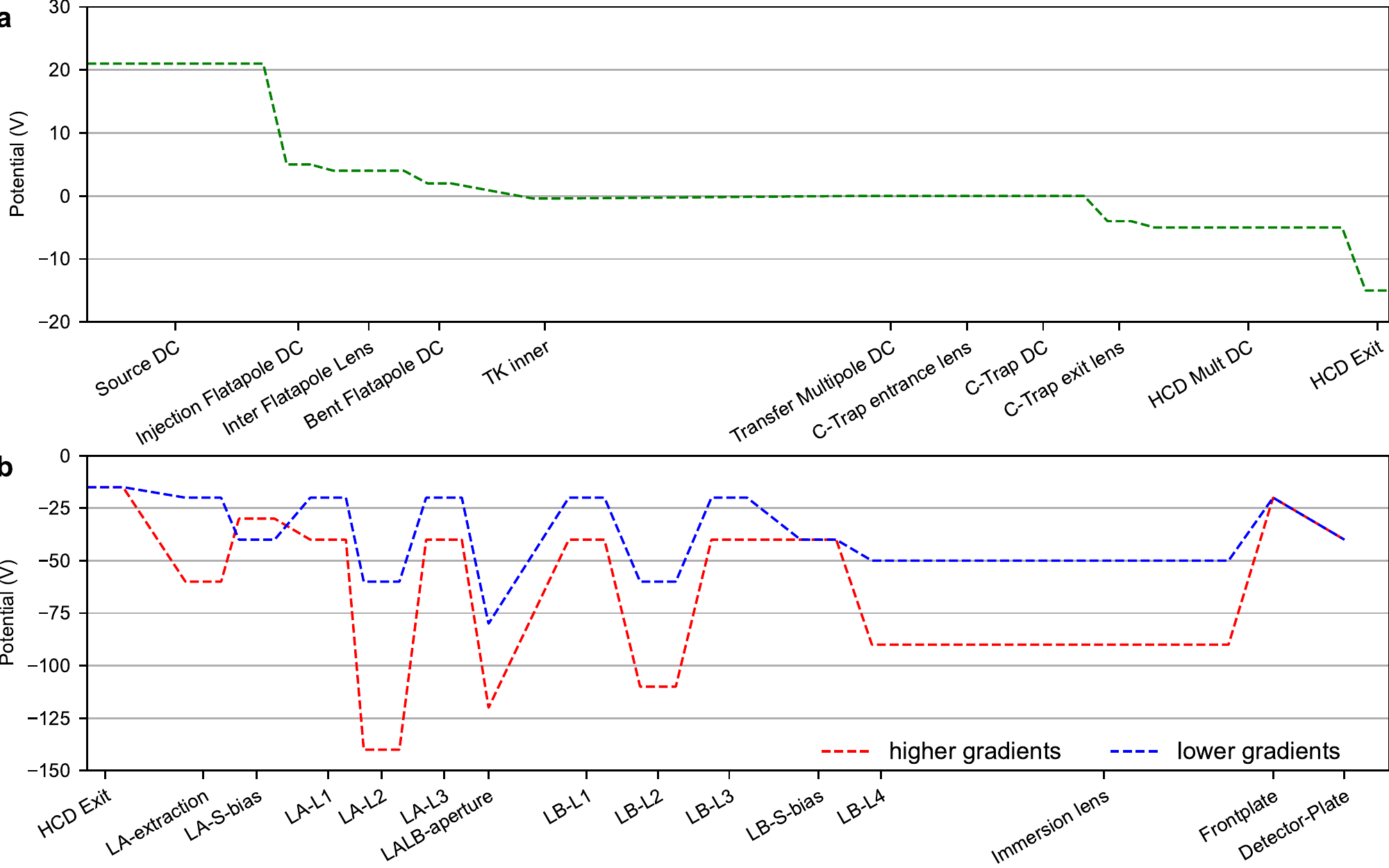}
 \caption{DC Potentials applied within \textbf{a} the mass spectrometer and \textbf{b} the custom landing stage for higher and lower DC gradients.}
 \label{fgr:UHMR_landing_potentials_SI}
\end{figure}

%Long version for SI
The ion beam is thermalised in the HCD cell at approx. $10^{-2}$ \si{mbar}.
% (calculated from the gas throughput at the QMS turbo pump, see labbok 5 p. 36 (in PDF))
The total ion beam energy (TE) is close to the effective potential therein (\SI{-5}{\electronvolt z^-1}). When the ions leave the HCD cell, they are accelerated in the electrostatic lens. In there, the pressure decreases from high $10^{-3}$ \si{mbar} (HCD side) to $10^{-6}$ \si{mbar} (deposition chamber side). As the electrostatic lens is longer (\SI{60}{mm} for high-pressure part)
% 57 mm from LA extraction to LALB Aperture, so roughly 60 mm, for pressure calculation look at labbook 7 p. 566
than BSA mean free path (\SI{0.1}{mm} for a native $BSA^{+14}$ ion at \SI{7e-3}{mbar}), collisions with the background gas occur.
% Ion trajectory simulations
% \todo[inline, author = Paul]{Include Tim's simulations?}
% \todo[inline, author = Tim]{As Stephan mentioned the simulation is rather a handy tool to make a figure demonstrating the effect than to argue that we have "discovered" it. I'm not sure if such a figure is even needed. Stephan?}
 % proof collisions cause the $E_{tot}$ to widen.
 For a hard-sphere collision, the kinetic energy E' of an ion after the collision is \cite{douglas_collisional_1992, douglas_mechanism_1982, cooks_collision_1978}:
 \begin{equation}
\label{eq:energy}
	\frac{E'}{E} = \frac{m_1^2 + m_2^2}{M^2} + \frac{2m_1 m_2}{M^2} \cdot cos(\theta_{cm})
\end{equation}
where $\theta_{cm}$ is the scattering angle in centre-of-mass coordinates, $m_1$ the ion mass, $m_2$ the gas molecule mass, $M = m_1 + m_2$ and $E$ the pre-collision ion kinetic energy. As $m_1 >> m_2$, the second term is close to zero. Thus, equation \ref{eq:energy} predicts $E'$ is a fraction of $E$ depending mostly on the ion and gas mass. Fig.~\ref{fgr:energy_loss_SI} compares this effect for heavy and light ions.
 The decrease in ion kinetic energy $E - E'$ is bigger for higher $E$. $E$ in the lab frame is proportional to the potential within the electrostatic lens. $E$ maximum is \SI{135}{\electronvolt z^-1} (strong gradient). As only \SI{8.8}{\electronvolt z^-1} (denatured) respectively \SI{5.2}{\electronvolt z^-1} (native) are dissipated in the electrostatic lens (Fig.~\ref{fgr:energy}), few high energy collisions occur.
For each $E'$ is close to the initial kinetic energy. Therefore, the $E_{tot}$ is lower after passing through stronger gradient conditions.

\begin{figure}
 \includegraphics[width=.9\textwidth]{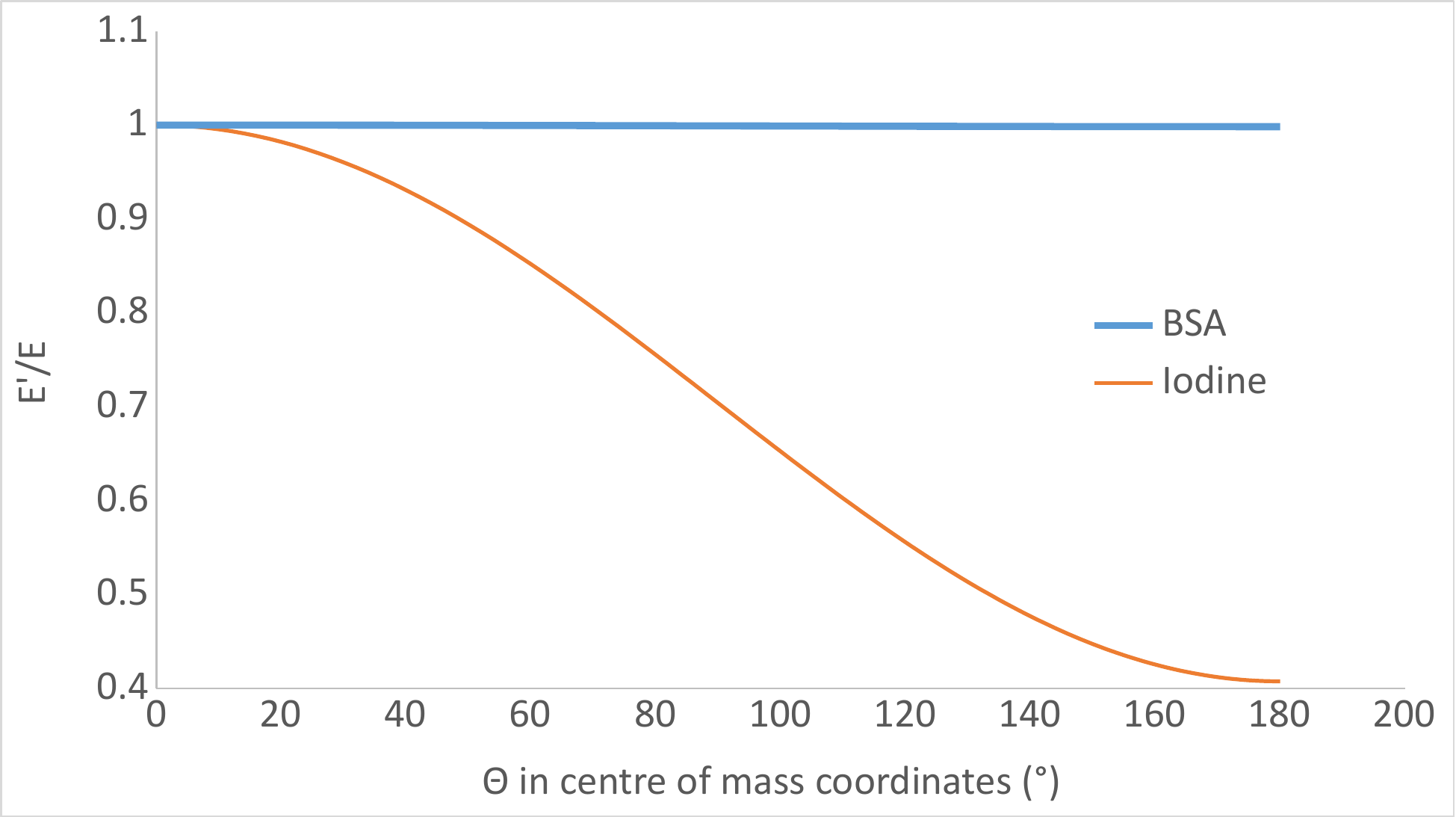}
 \caption{Kinetic energy loss per collision in \ce{N2} for BSA (mass \SI{66500}{\atomicmassunit}) and Iodine (mass \SI{127}{\atomicmassunit}) as a function of the scattering angle. Minimum for BSA is 0.99832 at \SI{180}{\degree}.}
 \label{fgr:energy_loss_SI}
\end{figure}

A similar argument applies to the distribution width: Arbitrary variations of the impact angle between a gas molecule and an ion cause variations in the scattering angle. Term 2 in equation \ref{eq:energy} varies accordingly and again $E'$ is a fraction of $E$, meaning absolute variations in ion kinetic energy $E - E'$ are bigger for higher $E$.

\begin{figure}
 \includegraphics[width=.9\textwidth]{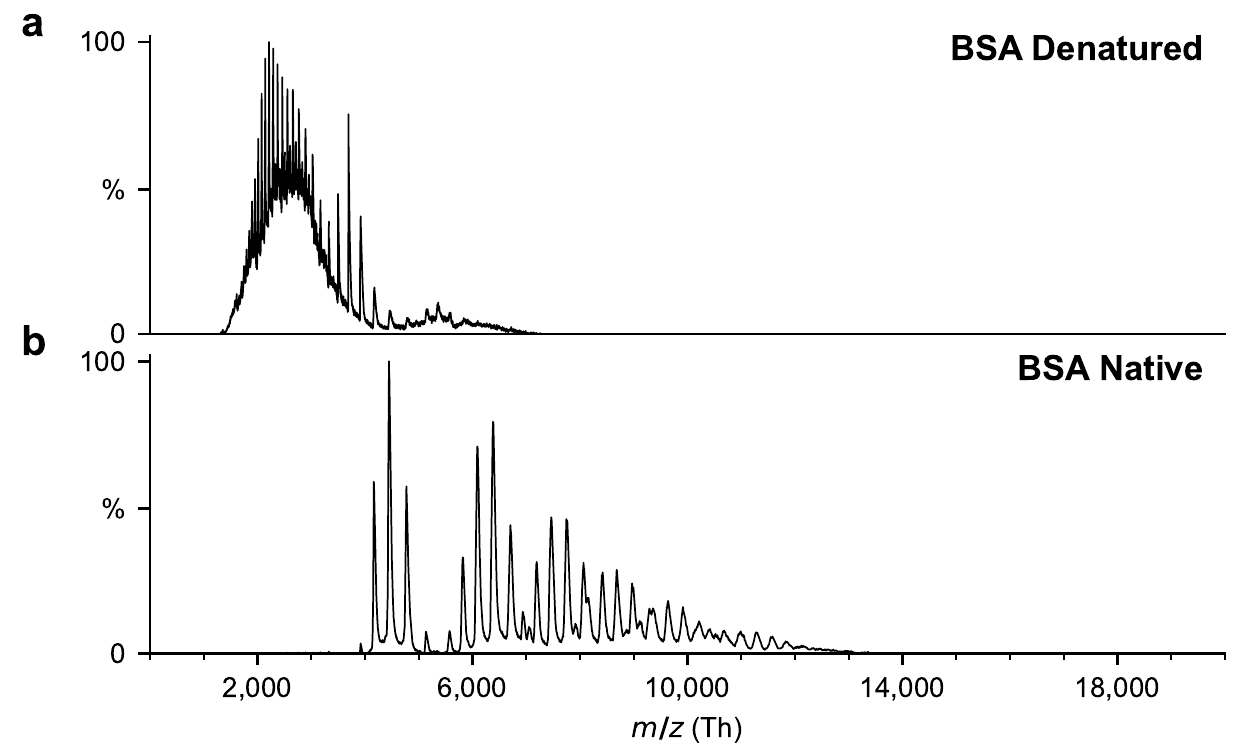}
 \caption{Mass spectra of \textbf{a} denatured BSA and \textbf{b} native BSA. Both mass spectra are acquired with non-activating conditions used for deposition.}
 \label{fgr:MS_BSA_SI}
\end{figure}

\newpage
\subsection{Transmission}
Native BSA emission current for the +15 charge state and a nano-ESI flow rate of \SI{1}{\ul h-{1}}
\begin{equation}
	I = \frac{cVzF}{t} = \frac{\SI{3e-6}{\mole\per\liter} \SI{1e-6}{l}\cdot 15 \cdot{\SI{96485}{C\per mol}}}{\SI{3600}{s}} = \SI{1.2}{nA}
\end{equation}
With concentration $c$, Volume $V$, number of charges $z$, Faraday constant $F$ and time $t$.

%Source for table: 2021 Desmond lab journal p. 134
\begin{table}
\centering
\begin{tabular}{llll}
\textbf{Measure at} & \textbf{$\Phi$ Optic (V)} & \textbf{Repulsive Optic (RO)} & \textbf{$\Phi$ RO (V)} \\
\hline
\hline
Emitter & 			$\approx$1200 & n.A. & n.A. \\
Transfer capillary & 21 & n.A. & n.A. \\
S exit lens & -100 & Injection flatapole DC & 50 \\
Inter flatapole lens & -50 & Bent flatapole & 50 \\
Inner TK lens & 0 & Outer TK & 60 \\
Aperture & -60 & steer beam & n.A. \\
Current detector & -20 & steer beam & n.A.
\end{tabular}
\caption{Potentials applied to ion optics when measuring current. Emitter potential is for nano-spray setup.}
\label{tab:SI_pot_current}
\end{table}

\newpage
\subsection{Ion beam size}
SIMION simulations are consistent with the observations in figure  \ref{fgr:spotsize}. However, since the angular velocity distribution of the beam upstream of the sample holder cannot be determined experimentally, a quantitative comparison of simulated and observed beam profiles is currently not meaningful.

\begin{figure}
 \includegraphics[width=.7\textwidth]{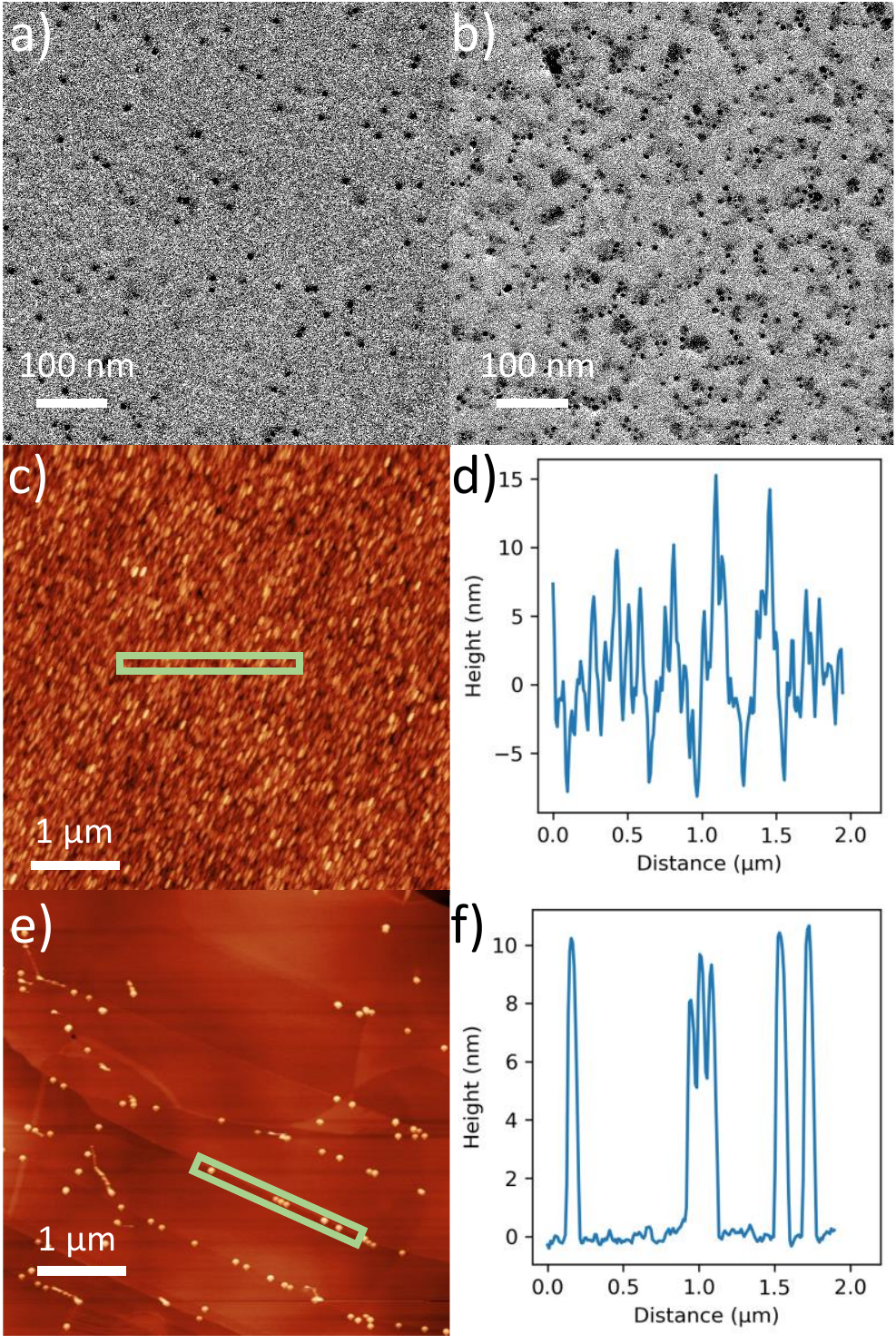}
 \caption{Exemplary AFM and TEM data for the beam shape characterization.
 \textbf{a)} and \textbf{b)} show low (250 particles per \textmu m$^2$) and high (1500 particles per \textmu m$^2$) density areas on the TEM grid.
 \textbf{c)} shows an area in the centre of the deposition spot with more than monolayer coverage. Therefore, the line profile \textbf{d)} shows no well-defined baseline.
 \textbf{e)} shows a less dense area with 4 particles per \textmu m$^2$. The corresponding line profile \textbf{f)} shows a clear baseline and allows to determine the particle heights.
 }
 \label{fgr:spotsize_SI}
\end{figure}

\newpage
\subsection{Mass filtering and solution composition}
\begin{figure}
 \includegraphics[width=0.5\textwidth]{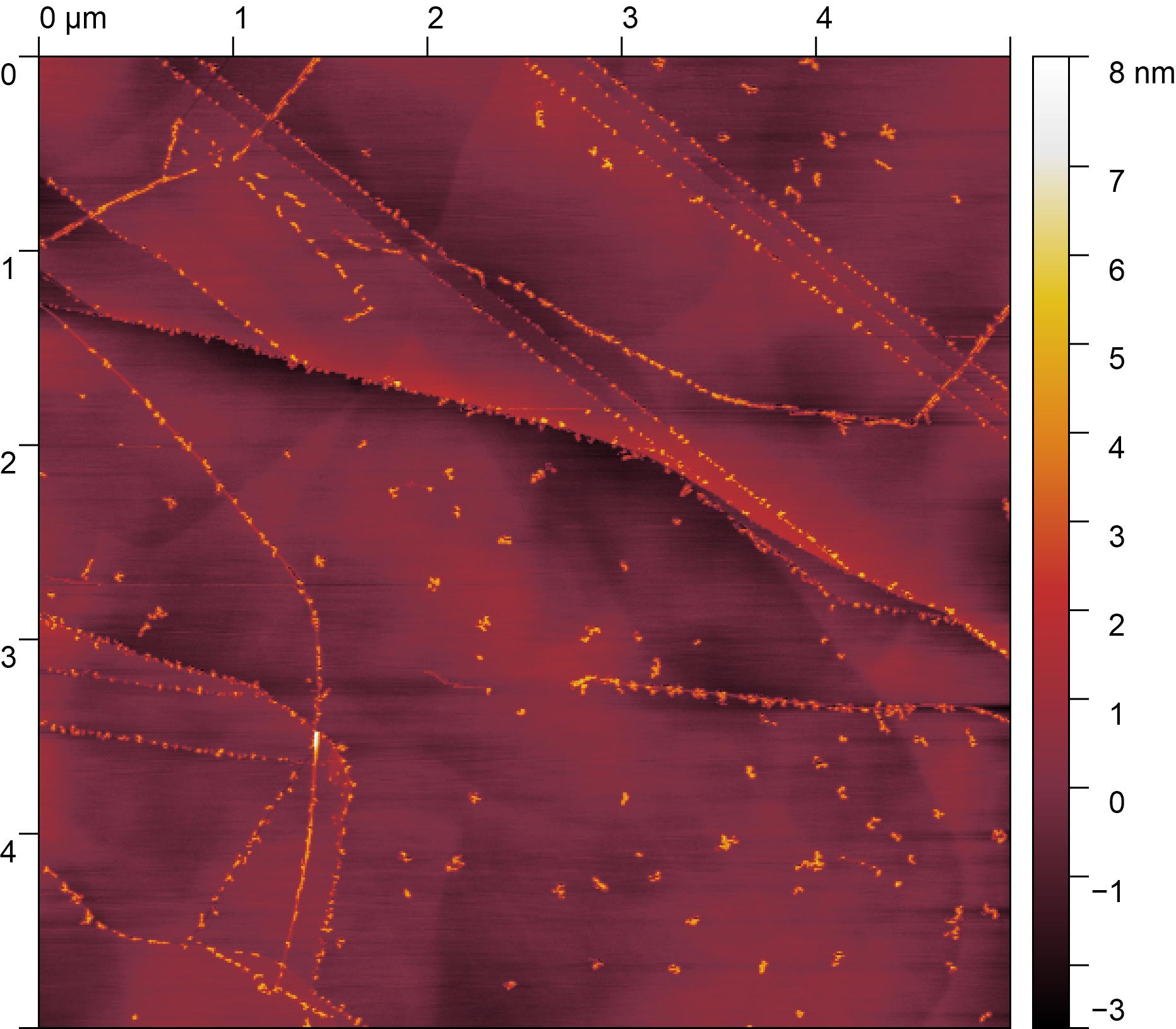}
 \caption{Native BSA on HOPG imaged with ambient AFM}
 \label{fgr:native_HOPG_SI}
\end{figure}

\begin{figure}
 \includegraphics[width=0.5\textwidth]{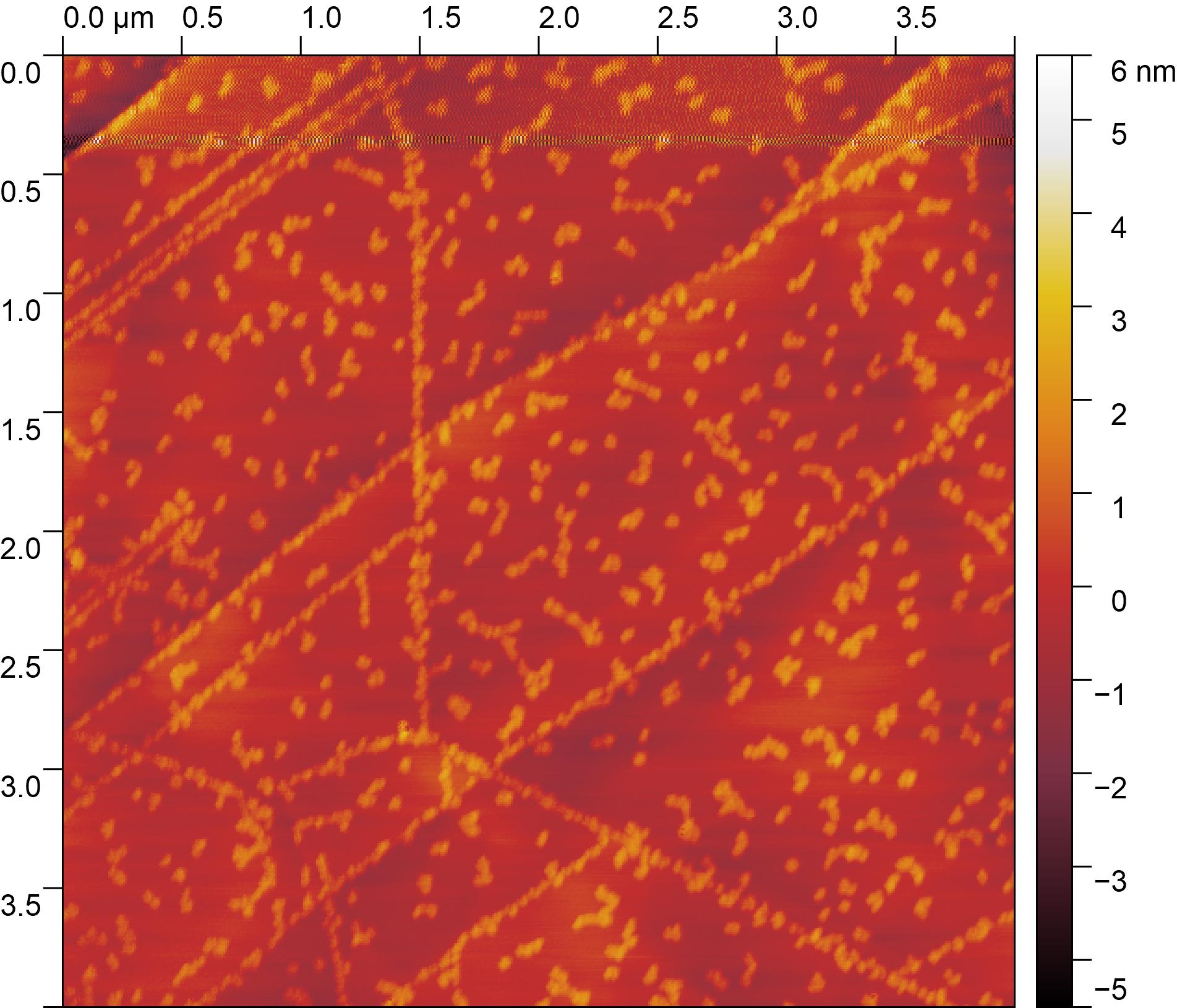}
 \caption{Denatured BSA on HOPG imaged with ambient AFM}
 \label{fgr:denatured_HOPG_SI}
\end{figure}

\newpage

\subsection{TEM}

\begin{figure}
 \includegraphics[width=.6\textwidth]{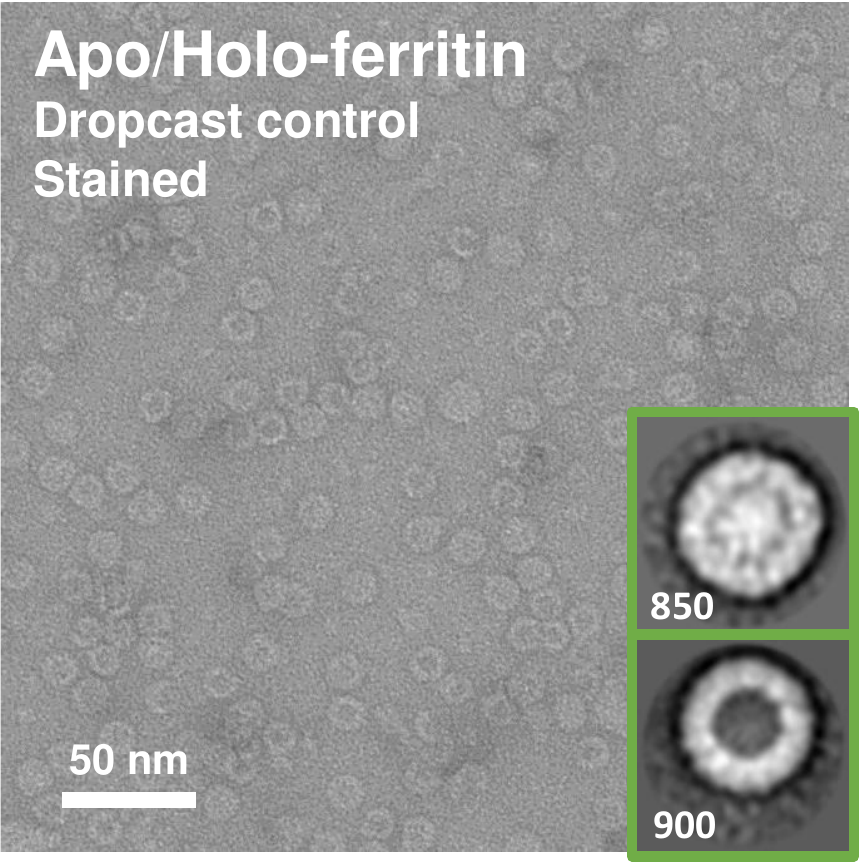}
 \caption{Negative stain apo/holo-ferritin control sample. The 2D classes show the same characteristic features as observed for the native ES-IBD samples, though they are better defined and there is less deformation in the control, despite the lower number of particles.
 }
 \label{fgr:TEM_SI}
\end{figure}

\newpage
\subsection{ADH activity}
\textbf{Method for extraction:} Before we established the successful protocol, we tried to extract the deposited ADH. All steps were the same as described for the submersion, except: We deposited two times \SI{27}{\ng} (\SI{128}{\pA h}) and two times \SI{37}{\ng} (\SI{175}{\pA h}) on four amorphous Carbon EM Grids (AGS160-4H, Agar Scientific Ltd, Stansted, Great Britain). Each \SI{27}{\ng} grid was left for 2 days after deposition in ambient conditions. Then, we put in \SI{50}{\ul} assay buffer in a PP micro centrifuge vial. One \SI{27}{\ng} grid was vortexed for \SI{15}{min}, the other sonicated for the same time. Then, the we transferred the extract in the 96 well plate.
For the \SI{37}{\ng} grids, both were transferred immediately after deposition in a well with \SI{50}{\ul} assay buffer. We moved them around for \SI{6}{min} with tweezers to wash ADH off. Then, we removed the grids.

% Unsuccessful but relevant previous experiments
\textbf{Results:} Neither sonication- nor vortex- nor washed-off-extracted ADH from EM-grids was active (Fig.~\ref{fgr:extraction_SI}). Submersion of the EM grid was not possible due to high background activity. We attribute this to a redox-reaction between the kit's components and the grid's copper support, which turned dull. Submersed conductive carbon tape was inert and used for all further work (Fig.~\ref{fgr:submersion_blanks_SI}).
To check if the adapted assay protocol was working, we used the nano-electrospray source to deposit two tapes at atmospheric pressure. Sample activity was \SI{2.1}{mU}, no background activity was present (Fig.~\ref{fgr:ESI_depo_SI}). Although charge can be measured, recovery calculation is not possible due to unknown composition of the ion-droplet-plume.

\begin{figure}
 \includegraphics[width=.85\textwidth]{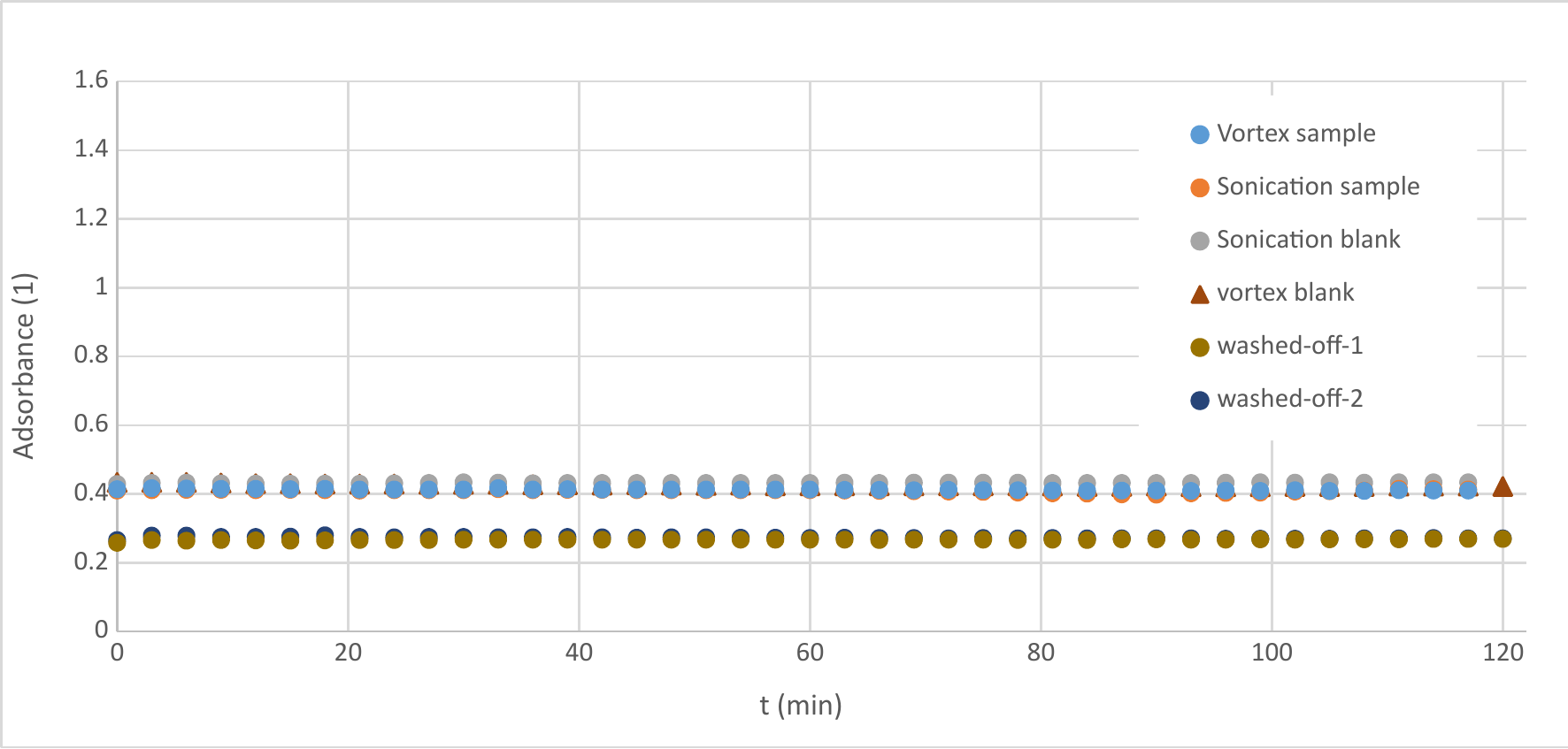}
 \caption{Assay-buffer-extraction of 128 pAh deposited ADH on EM grid with either vortex or sonication yields blank activity. The same applies for washing-off 175 pAh deposited ADH on EM grids. We tried to wash ADH off by moving the deposited grid around with tweezers in ADH buffer. 
 % \open{remove outer border}
 }
 \label{fgr:extraction_SI}
\end{figure}

\begin{figure}
 \includegraphics[width=.85\textwidth]{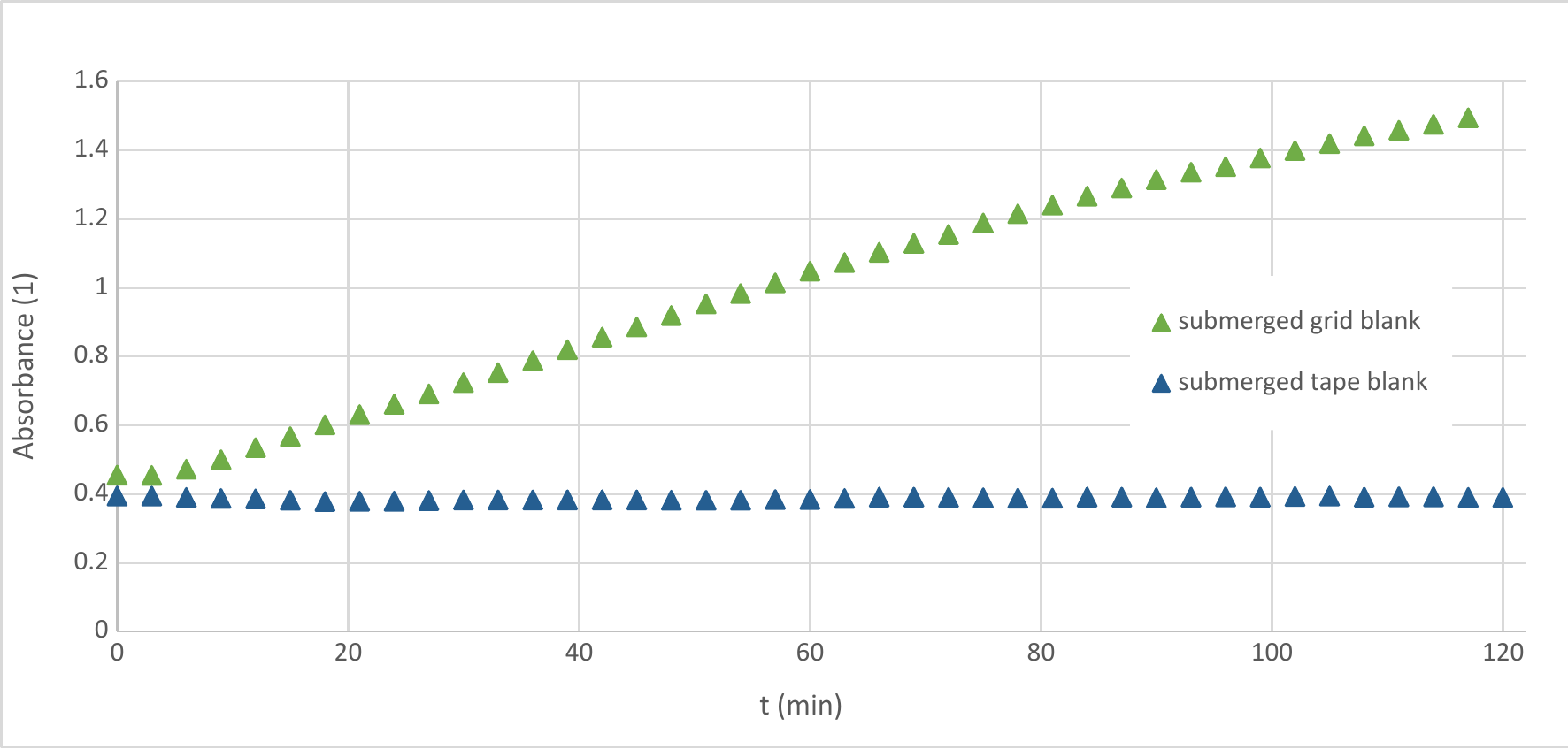}
 \caption{Submersion of an EM-grid in assay reaction mix causes a strong increase in absorbance, conductive tape doesn't.}
 \label{fgr:submersion_blanks_SI}
\end{figure}

\begin{figure}
 \includegraphics[width=0.48\linewidth]{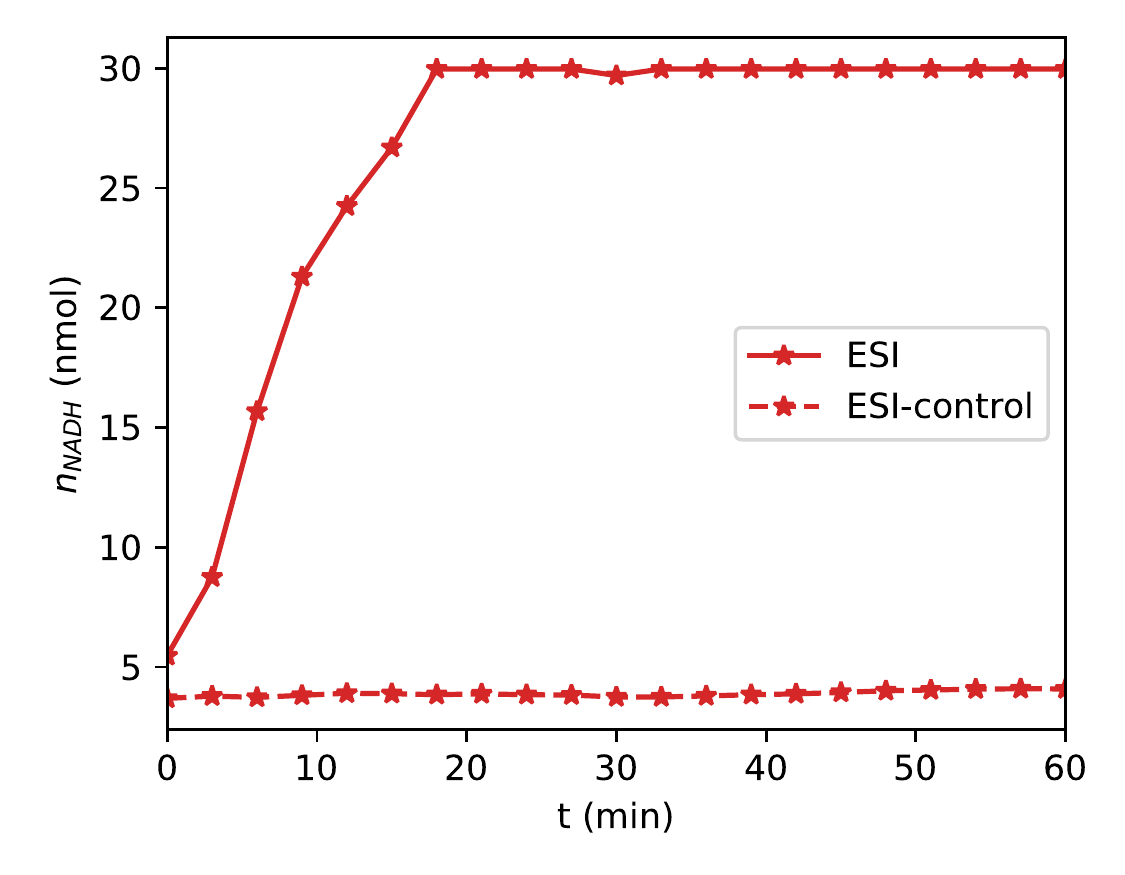}
 \caption{Production of NADH by ADH after electrospray deposition at atmospheric pressure. The broken lines stagnating at the offset level are background controls, so the NADH production is specific for ADH activity.}
 \label{fgr:ESI_depo_SI}
\end{figure}

\begin{figure}
 \includegraphics[width=.85\textwidth]{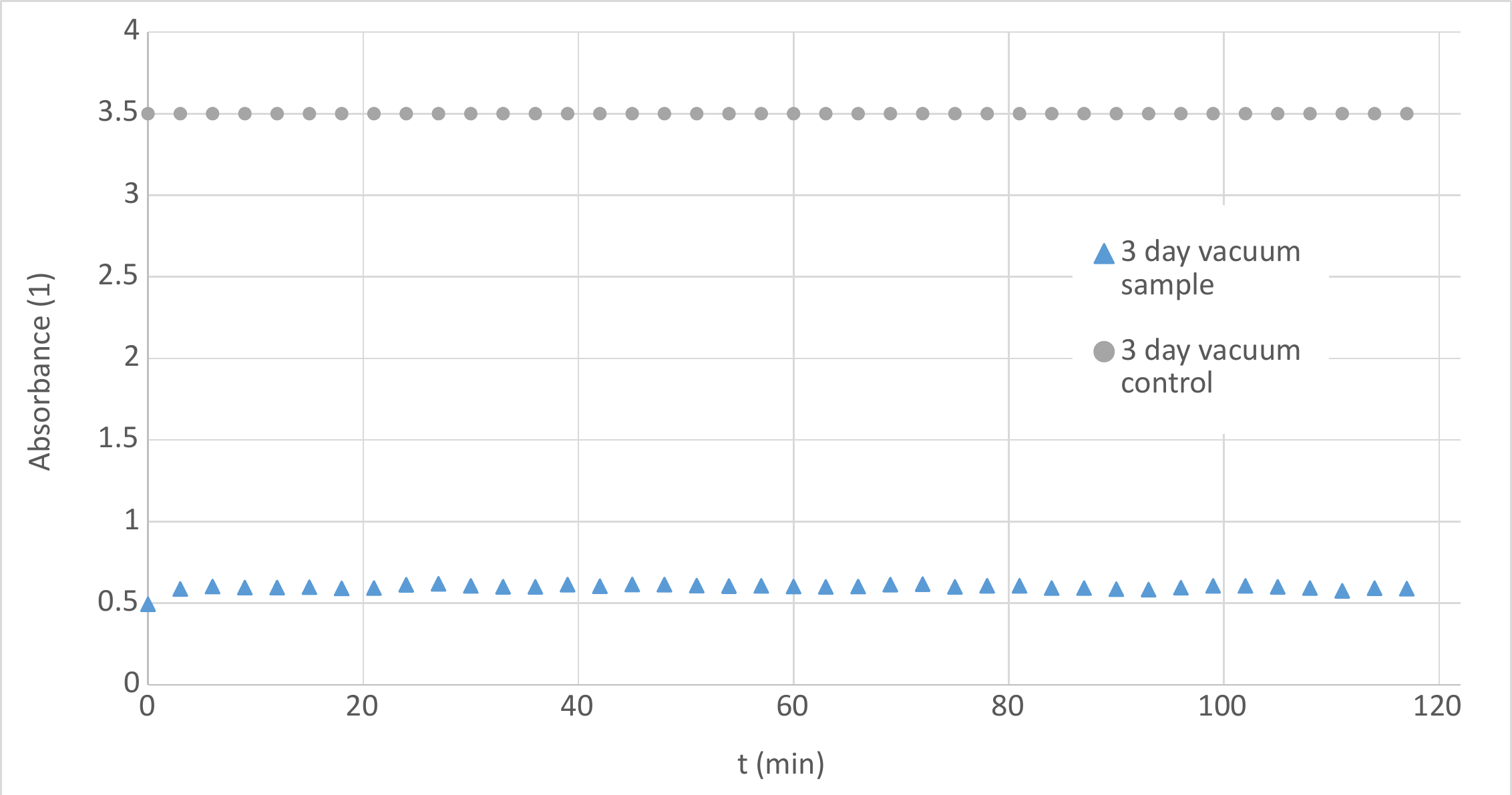}
 \caption{128 pAh deposition (repetition C) retains no activity after 3 day storage in vacuum. The corresponding control was inactive as well, but showed a high absorbance due to the conductive tape having moved into the beam path (see Fig.~\ref{fgr:blockage_SI}). 
 % \open{May be less confusing if we leave out this and the next fig and replace by a short statement. We were not able to demonstrate activity after 3 days but this was not reproduced.}
 }
 \label{fgr:storage_SI}
\end{figure}

\begin{figure}
 \includegraphics[width=.85\textwidth]{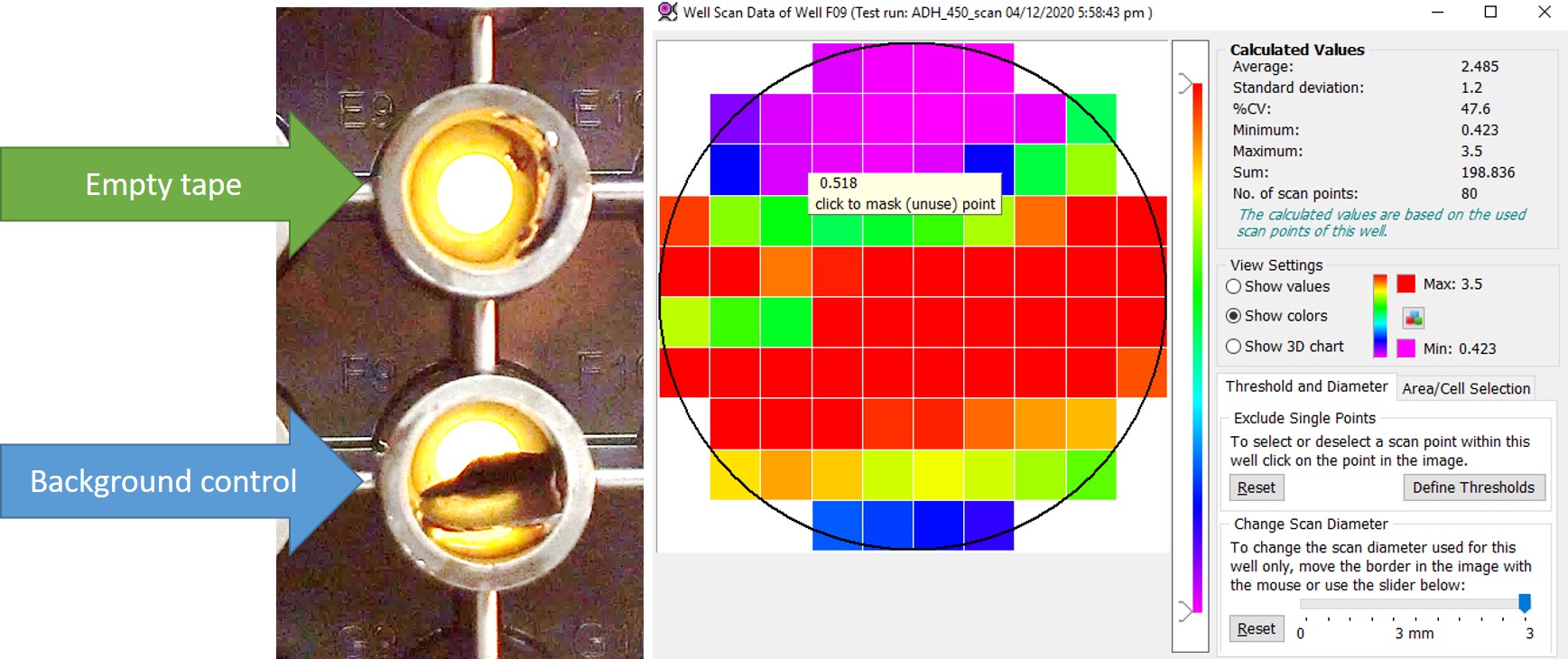}
 \caption{Left: 3 day storage background control (repetition C) moved into plate reader optical path. Note the liquid is still yellow, meaning no reaction has occurred during incubation. Right: 2D scan of bottom well confirms absorbance is still at blank level after incubation.}
 \label{fgr:blockage_SI}
\end{figure}

%The remaining activity can be calculated from the deposited amount:
%\begin{equation}
%	n_{ADH} = \frac{Q}{zF} = \frac{128 \cdot 10^{-12}A \cdot 3600s}{65 \cdot \frac{As}{mol}} = 1.8 \times 10^{-3} mol \hat{=} 2.7 \cdot 10^{-8} g
%\end{equation}
%\todo[inline]{Include calculation of yield for both the spray solution and the datasheet activity.}

\end{suppinfo}

\newpage
\bibliography{instBib} %instBibStephan

%% graphical table of contents.
%%%%%%%%%%%%%%%%%%%%%%%%%%%%%%%%%%%%%%%%%%%%%%%%%%%%%%%%%%%%%%%%%%%%%

% \begin{tocentry}

% \end{tocentry}

\end{document}